\documentclass[12pt,draftcls, onecolumn]{IEEEtran}

\usepackage{cite}

\usepackage{amssymb}
\ifCLASSINFOpdf
   \usepackage[pdftex]{graphicx}
\else
   \usepackage[dvips]{graphicx}
\fi

\usepackage[cmex10]{amsmath}
\usepackage{mdwmath}

\usepackage{array}

\usepackage{multirow}  
\usepackage{array} 
\newcommand{\PreserveBackslash}[1]{\let\temp=\\#1\let\\=\temp}
\newcolumntype{C}[1]{>{\PreserveBackslash\centering}p{#1}}
\newcolumntype{R}[1]{>{\PreserveBackslash\raggedleft}p{#1}}
\newcolumntype{L}[1]{>{\PreserveBackslash\raggedright}p{#1}}
\usepackage{chemarrow}

\ifCLASSOPTIONcompsoc 
\usepackage[caption=false,font=normalsize,labelfont=sf,textfont=sf]{subfig}
\else
\usepackage[caption=false,font=footnotesize]{subfig}
\fi

\usepackage{stfloats}
\usepackage{extarrows}

\begin{document}
\title{Fast DGT Based Receivers for GFDM in Broadband Channels}

\author{Peng~Wei, Xiang-Gen~Xia,~\IEEEmembership{Fellow, IEEE}, Yue~Xiao,~\IEEEmembership{Member, IEEE}, Shaoqian~Li,~\IEEEmembership{Fellow, IEEE} 
\thanks{P. Wei, Y. Xiao, and S. Li are with the National Key Laboratory of Science and Technology on Communications, University of Electronic Science and Technology of China, Chengdu, China (e-mail: pisces.wp@gmail.com; \{xiaoyue, lsq\}@uestc.edu.cn).}
\thanks{X.-G. Xia is with the Department of Electrical and Computer Engineering, University of Delaware, Newark, DE 19716 USA (e-mail: xianggen@udel.edu).}}

\maketitle

\begin{abstract}
\boldmath
Generalized frequency division multiplexing (GFDM) is a recent multicarrier 5G waveform candidate with flexibility of pulse shaping filters. However, the flexibility of choosing a pulse shaping filter may result in inter carrier interference (ICI) and inter symbol interference (ISI), which becomes more severe in a broadband channel. In order to eliminate the ISI and ICI, based on discrete Gabor transform (DGT), in this paper, a transmit GFDM signal is first treated as an inverse DGT (IDGT), and then a frequency-domain DGT is formulated to recover (as a receiver) the GFDM signal. Furthermore, to reduce the complexity, a suboptimal frequency-domain DGT called local DGT (LDGT) is developed. Some analyses are also given for the proposed DGT based receivers. 
\end{abstract}

\IEEEpeerreviewmaketitle

\section{Introduction}

\IEEEPARstart{G}{eneralized} frequency division multiplexing (GFDM) \cite{Ref1, Ref2, Ref12} has attracted much attention in recent years as a candidate waveform of 5G cellular systems for its low spectral leakage due to the flexibility of its pulse shaping filter \cite{Ref1, Ref2, Ref12, Ref8, Ref10, Ref19, Ref17,Ref18, Ref14, Ref15, Ref16}. A pulse shaping filter with better spectral property, however, may cause intersymbol interference (ISI) and inter carrier interference (ICI), which becomes more severe in a broadband channel and may cause problems at the receiver. Among the methods in \cite{Ref1, Ref2} for signal recovery in the receiver for a GFDM system, matched filter (MF) receiver maximizes the signal-to-noise ratio (SNR) while causing self-interference from the nonorthogonality of the transmit waveform. Zero-forcing (ZF) receiver can cancel the self-interference at the price of the channel noise enhancement. To reduce the high self-interference in MF, MF with successive interference cancellation (MF-SIC) receiver is presented in \cite{Ref2} at the cost of high-complexity iterative processing. Linear minimum mean square error (MMSE) receiver can improve the performance of ZF receiver. However, based on the transmitter matrix for generating the GFDM signal, these GFDM receivers have high complexities proportional to the square of the total number of the data symbols in a GFDM symbol. To obtain a low-complexity implementation in the GFDM receiver, based on fast Fourier transform (FFT) and inverse FFT (IFFT), FFT-based ZF/MF \cite{Ref1}, FFT-based MF-SIC \cite{Ref10} and several techniques for MF \cite{Ref18, Ref14, Ref15, Ref16}, ZF \cite{Ref18, Ref15, Ref16}, and MMSE \cite{Ref17, Ref15, Ref16} are proposed. In the ideal channel, among the low-complexity methods, the ZF/MF receiver in \cite{Ref15, Ref16} can obtain the lowest complexity by splitting the multiplication of the transmitter matrix and discrete Fourier transform (DFT)/inverse DFT (IDFT) matrix into small blocks with FFT/IFFT implementation. In a broadband channel, besides the complexity of the techniques themselves, another key factor is the channel equalization that should be considered in the receiver. Since the direct channel equalization in time domain in \cite{Ref1} has a high complexity proportional to the square of the total number of the data symbols in a GFDM symbol, frequency domain equalization (FDE) can be used to reduce the complexity \cite{Ref15, Ref16}. In this case, the proposed receivers in \cite{Ref15, Ref16} have lower computational cost than the low-complexity receivers in \cite{Ref1}. Unfortunately, compared to the orthogonal frequency multiplexing division (OFDM) receiver, the FDE \cite{Ref15, Ref16} needs extra FFT/IFFT operations, where in \cite{Ref15, Ref16}, it is called ZF/MF receiver directly and its complexity will be compared in details. 

In this paper, to simplify the GFDM receiver for a broadband channel similar to the OFDM receiver, a relationship between a GFDM signal and discrete Gabor transform (DGT) \cite{Ref3, Ref4, Ref13} is first investigated, similar to \cite{Ref8}, \cite{Ref19}, i.e, a transmit GFDM signal is an inverse DGT (IDGT) of a data array. Then, according to DGT \cite{Ref3, Ref4, Ref13}, a frequency-domain DGT is proposed for GFDM signal recovery, which is different from the time-domain DGT in \cite{Ref8}, \cite{Ref19} causing high-complexity time-domain channel equalization. By analyzing the interference after the frequency-domain DGT for GFDM signals, we conclude that the coherence bandwidth, related to the reciprocal of the maximum channel delay, and the roll-off factor of a transmit waveform are two key factors of the interference in a GFDM system, where high coherence bandwidth and small roll-off factor can make the GFDM signal recovered by the frequency-domain DGT much like OFDM. Furthermore, to reduce the complexity of the frequency-domain DGT in the whole band, a suboptimal frequency-domain DGT in local subbands, called local DGT (LDGT), is proposed. 
Simulation results show that the frequency-domain DGT with small roll-off factor can achieve considerable bit-to-error rate (BER) performance close to OFDM, and LDGT significantly reduces the complexity of the frequency-domain DGT with a small BER performance degradation.

The rest of the paper is organized as follows. In Section II, GFDM signals are formulated in transmitter as IDGT and in receiver as DGT,  and the frequency domain DGT is proposed. In Section III, a received GFDM signal is formulated by the frequency-domain DGT followed by analyzing the interference generated in the frequency-domain DGT, and LDGT is presented and analyzed for complexity reduction. In Section IV, simulation results for the frequency-domain DGT, LDGT, and several other existing GFDM signal recovery methods are presented. Finally, in Section V, this paper is concluded.  

\section{GFDM, DGT, IDGT, and Frequency-Domain DGT}

In this section, transmitted and received GFDM signals are first briefly introduced. Then, based on the theory of DGT, an IDGT is investigated for a transmitted GFDM signal. Lastly, a frequency-domain DGT is proposed for the GFDM signal recovery.
\subsection{GFDM Signal}

In GFDM transmitter, bit streams are first modulated to complex symbols $d_{k, m}$ that are divided into sequences of \emph{KM} symbols long. Each sequence (as a vector) $\mathbf{d}=[\mathbf{d}^{T}_{0}, \mathbf{d}^{T}_{1}, \ldots, \mathbf{d}^{T}_{M-1}]^{T}$ with $\mathbf{d}_m=[d_{0,m}, d_{1,m}, \ldots, d_{K-1,m}]^{T}$, $m=0, 1, \ldots, M-1$, is spread on \emph{K} subcarriers in \emph{M} time slots. Therein, $d_{k,m}$ is the transmitted data on the \emph{k}th subcarrier in the \emph{m}th subsymbol of each GFDM block. The data symbols are taken from a zero mean independent and identically distributed (i.i.d) process with the unit variance. Each $d_{k,m}$ is transmitted with a pulse shaping filter \cite{Ref1}
\begin{equation}
\label{Eqn1}
g_{k,m}(n)=g\left((n-mK)_N\right)e^{-j2\pi\frac{k}{K}n},
\end{equation}
where the signal sample index is $n=0,1,\ldots, N-1$ with $N=KM$ satisfying the condition of critical sampling in DGT, $(\cdot)_N$ denotes the modulo of \emph{N}, and $g(n)$ is a prototype filter whose time and frequency shifts are $g_{k,m}(n)$. By the superposition of all the filtered $d_{k,m}$, the GFDM signal in transmission is
\begin{equation}
\label{Eqn2}
x(n)=\sum\limits^{K-1}_{k=0}{\sum\limits^{M-1}_{m=0}{d_{k,m}g_{k,m}(n)}}.
\end{equation}

At the receiver, the received GFDM signal is
\begin{equation}
\label{Eqn3}
y(n)=h(n)* x(n)+w(n),
\end{equation}
where $*$ denotes the linear convolution operation, $h(n)$ is the channel response in the time domain, and $w(n)$ is the AWGN noise with zero mean and variance $\sigma^2$. 

\begin{figure}[ht]
\centering
\includegraphics[width=6in]{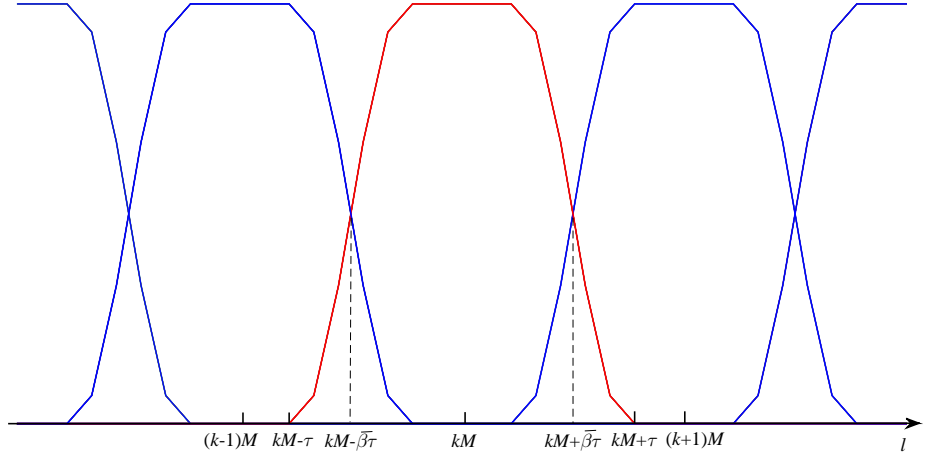}
\caption{Frequency-domain GFDM transmitting filter $G_{k,m}(l)$ where $\bar{\beta}=1/(1+\beta)$ and $\beta$ is the roll-off factor of $g(n)$.}
\label{fig:fig1}
\end{figure}

Assuming perfect synchronization and long enough cyclic prefix (CP) against the maximum channel delay are implemented, the frequency-domain expression of \eqref{Eqn3} can be written as
\begin{equation}
\label{Eqn4}
Y(l)=H(l)X(l)+W(l),
\end{equation}
where $l=0,1,\ldots,N-1$, $X(l)$ is the \emph{N}-point DFT of $x(n)$ as
\begin{equation}
\label{Eqn5}
X(l)=\sum\limits^{K-1}_{k=0}{\sum\limits^{M-1}_{m=0}{d_{k,m}G_{k,m}(l)}},
\end{equation}
and $G_{k,m}(l)$ is the \emph{N}-point DFT of $g_{k,m}(n)$ as
\begin{align}
 \label{Eqn7}
G_{k,m}(l)
    &  =  \sum\limits^{N-1}_{n=0}{g_{k,m}(n)e^{-j2\pi\frac{l}{N}n}}  \nonumber \\
    & =\sum\limits^{N-1}_{n=0}{g((n-mK)_N)e^{-j2\pi\frac{l+kM}{N}n}} \nonumber \\
    & \stackrel{n^{\prime}=n-mK}{=}\sum\limits^{N-1-mK}_{n^{\prime}=-mK}{g((n^{\prime})_N)e^{-j2\pi\frac{l+kM}{N}(n^{\prime}+mK)}} \nonumber \\
    & =e^{-j2\pi\frac{l}{M}m}e^{-j2\pi km}\sum\limits^{N-1-mK}_{n^{\prime}=-mK}{g((n^{\prime})_N)e^{-j2\pi\frac{l+kM}{N}n^{\prime}}} \nonumber \\
    & =e^{-j2\pi\frac{l}{M}m}\sum\limits^{N-1}_{n=0}{g(n)e^{-j2\pi\frac{l+kM}{N}n}} \nonumber \\
    & =G((l+kM)_N)e^{-j2\pi\frac{l}{M}m},
\end{align}
where $G(l)$ for $l=0, 1, \ldots, N-1$ is the \emph{N}-point DFT of $g(n)$ for $n=0, 1, \ldots, N-1$, and thus the frequency and time shifts of $G(l)$ are $G_{k,m}(l)$ shown in Fig. \ref{fig:fig1}. In Fig. \ref{fig:fig1}, 
\begin{equation}
\label{Eqn8}
G((l)_N)=\begin{cases}
    f(l), & -\tau\leqslant l\leqslant\tau, \\
    0,  & \text{otherwise},
\end{cases}
\end{equation}
where $f(l)$ is a baseband-equivalent window function in the frequency domain, for example, the raised cosine (RC) function, the root raised cosine (RRC) function and the Xia pulse \cite{Ref6}, integer \emph{l} is in the finite interval $[-N/2, N/2-1]$, and $\tau$ is a positive integer satisfying $\tau \leqslant N/2$ and denotes the window width.  
Additionally, the local property of $G(l)$ can save the storage compared to the $N\times N$ transmitter matrix in \cite{Ref1}.

To demodulate the GFDM signal after the time-domain channel equalization, MF, ZF, linear MMSE, and MF-SIC receivers are proposed in \cite{Ref1}. However, when the transmitter matrix has a large size, these receivers with the time-domain channel equalization have high complexities. Our goal here is to simplify the GFDM receiver with insignificant ICI and ISI.  

\subsection{DGT, IDGT, and Frequency-Domain DGT}

Without the channel influence, i.e., in an ideal channel, in order to cancel the ISI and ICI for the GFDM signal recovery, the properties of the transmitted GFDM signal should be first investigated. To do so, let us briefly review DGT and IDGT. 

For a signal $x(n)$, $n=0,1,...,N-1$, its DGT is defined as 
\begin{equation}
\label{Eqn10}
d_{k,m}=\sum\limits^{N-1}_{n=0}{x(n)\gamma^{*}_{k,m}(n)},
\end{equation}
where the time and frequency shifts of an analysis window $\gamma(n)$ are
\begin{equation}
\label{Eqn11}
\gamma_{k,m}(n)=\gamma\left((n-mK)_N\right)e^{-j2\pi\frac{kM}{N}n}.
\end{equation}

The IDGT of $d_{k,m}$ are defined as
\begin{equation}
\label{Eqn9}
x(n)=\sum\limits^{K-1}_{k=0}{\sum\limits^{M-1}_{m=0}{d_{k,m}g\left((n-mK)_N\right)e^{-j2\pi\frac{k}{K}n}}},
\end{equation}
where $g(n)$ is a synthesis window, which is the same as the GFDM transmitted signal in \eqref{Eqn2}.  
When $g(n)$ and $\gamma(n)$ satisfy the following Wexler-Raz identity, $d_{k, m}$ and $x(n)$ in \eqref{Eqn10} and \eqref{Eqn9} are the same:
\begin{align}
\label{Eqn12}
& \sum\limits^{N-1}_{n=0} {g(n+kK)e^{-j2\pi \frac{mM}{N}n}\gamma^{*}(n)}=\delta(k)\delta(m) \nonumber \\
&\qquad 0\leqslant k\leqslant M-1, \quad \text{and} \quad  0\leqslant m\leqslant K-1.
\end{align}
In this case, DGT is the receiver while IDGT is the transmitter, and \eqref{Eqn10} and \eqref{Eqn9} form a pair.


Furthermore, from \eqref{Eqn5} and \eqref{Eqn9}, \eqref{Eqn5} is the IDGT of $d_{k,m}$ in the frequency domain. Thus, from \eqref{Eqn10}, the frequency domain DGT, as a pair with the frequency domain IDGT in \eqref{Eqn5}, is 
\begin{equation}
\label{Eqn16}
d_{k,m}=\frac{1}{N}\sum\limits^{N-1}_{l=0}{X(l){\Gamma}^{*}_{k,m}(l)},
\end{equation}
where $1/N$ is from the \emph{N}-point IDFT and 
\begin{align}
 \label{Eqn15}
\Gamma_{k,m}(l)
     =\Gamma((l+kM)_N)e^{-j2\pi\frac{m}{M}l},
\end{align}
which are the frequency and time shifts of $\Gamma(l)$ for $l=0,1,\ldots,N-1$, and $\Gamma(l)$ is the \emph{N}-point DFT of $\gamma(n)$. According to the Wexler-Raz identity \cite{Ref3, Ref4, Ref13}, the biorthogonality between the synthesis window $G(l)$ and the analysis window $\Gamma(l)$ is expressed by
\begin{align}
\label{Eqn17}
& \frac{1}{N}\sum\limits^{N-1}_{l=0} {G((l+mM)_N)e^{j2\pi \frac{k}{M}l}\Gamma^{*}(l)}=\delta(k)\delta(m) \nonumber \\
&\qquad 0\leqslant k\leqslant M-1, \; \text{and} \;  0\leqslant m\leqslant K-1.
\end{align}



In summary, for a GFDM signal over an ideal channel, it can be recovered by its DGT in either time domain \eqref{Eqn10} or frequency domain \eqref{Eqn16}. In other words, \eqref{Eqn10} or \eqref{Eqn16} is a receiver for GFDM signals in an ideal channel or a narrow band channel. The reason why the frequency domain DGT is mentioned here is for a broadband channel in next section. 

\section{Frequency-Domain DGT Receiver for GFDM Signals over a Broadband Channel}

In this section, we formulate a received GFDM signal similar to OFDM by the proposed frequency-domain DGT in a broadband channel. Two models are proposed and analyzed, in which the frequency-domain DGT in the whole band is considered in the first model and LDGT is proposed in the second model for the complexity reduction.
\subsection{Frequency-Domain DGT Model in the Whole Band}

From \eqref{Eqn3}, in a broadband channel, to use the time-domain DGT at the receiver, the time-domain channel equalization in the whole GFDM symbol of length \emph{N} has a high complexity, i.e., $O(N^2)$, and the FDE of the channel needs a pair of  \emph{N}-point FFT and \emph{N}-point IFFT. In contrast, as we shall see below, after \emph{N}-point FFT, the frequency-domain DGT can be adopted for the GFDM signal recovery, where the channel equalization has much lower complexity than the time-domain equalization and reduces an IFFT compared to FDE. Moreover, after the frequency-domain DGT for the coded GFDM signal, without a direct channel equalization, the signal with the channel information can be directly used to calculate the soft information for the decoder.

Substituting \eqref{Eqn4} into \eqref{Eqn16}, the frequency-domain DGT of the received GFDM signal $Y(l)$ in the broadband channel is expressed by
\begin{align}
\label{Eqn18}
Y_{k,m}&=\frac{1}{N}\sum\limits^{N-1}_{l=0}{Y(l){\Gamma}^{*}_{k,m}(l)}  \nonumber\\
&=\frac{1}{N}\sum\limits^{N-1}_{l=0}{H(l)X(l){\Gamma}^{*}_{k,m}(l)}+\frac{1}{N}\sum\limits^{N-1}_{l=0}{W(l){\Gamma}^{*}_{k,m}(l)}  \nonumber\\
&=\frac{1}{N}\sum\limits^{N-1}_{l=0}{H(kM)X(l){\Gamma}^{*}_{k,m}(l)}+\frac{1}{N}\sum\limits^{N-1}_{l=0}{(H(l)-H(kM))X(l){\Gamma}^{*}_{k,m}(l)} +\frac{1}{N}\sum\limits^{N-1}_{l=0}{W(l){\Gamma}^{*}_{k,m}(l)}  \nonumber\\
&=H(kM)d_{k,m}+\Omega_{k,m}+\Psi_{k,m},
\end{align}
where $H(kM)$ is the frequency-domain channel response corresponding to the \emph{k}th subcarrier,
\begin{align}
\label{Eqn19}
\Omega_{k,m}&=\frac{1}{N}\sum\limits^{N-1}_{l=0}{(H(l)-H(kM))X(l){\Gamma}^{*}_{k,m}(l)}  \nonumber \\ 
&=\frac{1}{N}\sum\limits^{N-1}_{l=0}{(H(l)-H(kM))X(l)\Gamma^{*}((l+kM)_N)e^{j2\pi\frac{m}{M}l}},
\end{align}
and 
\begin{equation}
\label{Eqn20}
\Psi_{k,m}=\frac{1}{N}\sum\limits^{N-1}_{l=0}{W(l)\Gamma^{*}((l+kM)_N)e^{j2\pi\frac{m}{M}l}}.
\end{equation}
It is shown in \eqref{Eqn18} that after the frequency-domain DGT, the (\emph{k}, \emph{m})-th GFDM symbol $Y_{k,m}$ has the similar format to the received OFDM symbol in the frequency domain. Then, the symbol-by-symbol detection is
\begin{equation}
\label{Eqn21}
\hat{d}_{k,m}=\arg\min\limits_{d_{k,m}\in\mathcal{S}}{\left|{Y}_{k,m}-H(kM)d_{k,m}\right|^2},
\end{equation}
where $\mathcal{S}$ is the signal constellation.

From \eqref{Eqn18} one can see that the received signal is corrupted by the interference $\Omega_{k,m}$ and the channel noise $\Psi_{k,m}$. For the frequency-domain DGT, the distortion composed of $\Omega_{k,m}$ and $\Psi_{k,m}$ is different from the Gaussian noise in OFDM systems. Since the Gaussian noise part $\Psi_{k,m}$ can be studied easily and similarly to before, we focus our analysis on the interference $\Omega_{k,m}$. It is shown in \eqref{Eqn19} that $\Omega_{k,m}$ is affected by the channel response $H(l)$ and the shifted analysis window ${\Gamma}_{k,m}(l)$, which will be analyzed in the following. 

Assuming $E\{\mathbf{d}\mathbf{d}^{H}\}=\mathbf{I}_N$ with the identity matrix $\mathbf{I}_N$ and $\mathbf{d}=[d_{0,0}, \ldots, d_{K-1,0}, d_{0,1}, \ldots, d_{K-1,M-1}]^T$, the variance of $\Omega_{k,m}$ can be expressed by
\begin{align}
& E\left\{\Omega^{*}_{k,m}\Omega_{k,m}\right\}  \nonumber \\
& =E\left\{\frac{1}{N}\sum\limits^{N-1}_{l=0}{(H(l)-H(kM))^*X^*(l){\Gamma}_{k,m}(l)}\frac{1}{N}\sum\limits^{N-1}_{\bar{l}=0}{(H(\bar{l})-H(kM))X(\bar{l}){\Gamma}^*_{k,m}(\bar{l})}\right\} \nonumber \\
& =\frac{1}{N^2}\sum\limits^{N-1}_{l=0} {\sum\limits^{N-1}_{\bar{l}=0}{E\left\{(H(l)-H(kM))^*(H(\bar{l})-H(kM))\right\}E\left\{X^*(l)X(\bar{l})\right\}}{\Gamma}_{k,m}(l){\Gamma}^*_{k,m}(\bar{l})}   \nonumber 
\end{align}
where $E\{\cdot\}$ denotes the expectation.  
Suppose that $N_{\rm c}$ paths are in the Jakes' model \cite{Ref11} of the Rayleigh fading channel with the discrete maximum Doppler shift $k_{D}$, the index set of the paths is $\mathcal{N}$, and $P_{\rm h}$ is the average power per path in the fading channel. Then, we can obtain
\begin{align} 
\label{Eqn22}
& E\left\{(H(l)-H(kM))^*(H(\bar{l})-H(kM))\right\} \nonumber \\
& = \sum\limits_{n_{\rm c}\in \mathcal{N}}{\sum\limits_{\bar{n}_{\rm c}\in \mathcal{N}}{E\left\{h^*(n_{\rm c})h(\bar{n}_{\rm c})\right\}e^{j2\pi \frac{l}{N}n_{\rm c}}e^{-j2\pi \frac{\bar{l}}{N}\bar{n}_{\rm c}}}}-\sum\limits_{n_{\rm c}\in \mathcal{N}}{\sum\limits_{\bar{n}_{\rm c}\in \mathcal{N}}{E\left\{h^*(n_{\rm c})h(\bar{n}_{\rm c})\right\}e^{j2\pi \frac{l}{N}n_{\rm c}}e^{-j2\pi \frac{kM}{N}\bar{n}_{\rm c}}}}  \nonumber \\
& \quad -\sum\limits_{n_{\rm c}\in \mathcal{N}}{\sum\limits_{\bar{n}_{\rm c}\in \mathcal{N}}{E\left\{h^*(n_{\rm c})h(\bar{n}_{\rm c})\right\}e^{j2\pi \frac{kM}{N}n_{\rm c}}e^{-j2\pi \frac{\bar{l}}{N}\bar{n}_{\rm c}}}}+\sum\limits_{n_{\rm c}\in \mathcal{N}}{\sum\limits_{\bar{n}_{\rm c}\in \mathcal{N}}{E\left\{h^*(n_{\rm c})h(\bar{n}_{\rm c})\right\}e^{j2\pi \frac{kM}{N}n_{\rm c}}e^{-j2\pi \frac{kM}{N}\bar{n}_{\rm c}}}} \nonumber \\
& = \sum\limits_{n_{\rm c}\in \mathcal{N}}{\sum\limits_{\bar{n}_{\rm c}\in \mathcal{N}}{P_{\rm h} J_0\left(2\pi \frac{k_{D}}{N}(n_{\rm c}-\bar{n}_{\rm c})\right)\left(e^{j2\pi \frac{l}{N}n_{\rm c}}-e^{j2\pi \frac{kM}{N}n_{\rm c}}\right)\left(e^{-j2\pi \frac{\bar{l}}{N}\bar{n}_{\rm c}}-e^{-j2\pi \frac{kM}{N}\bar{n}_{\rm c}}\right)}}  \nonumber \\
& = \sum\limits_{n_{\rm c}\in \mathcal{N}}{\sum\limits_{\bar{n}_{\rm c}\in \mathcal{N}}{P_{\rm h} \sum\limits^{\infty}_{s=0}{\frac{(-1)^s}{s!s!}\left(\frac{2\pi k_{D}(n_{\rm c}-\bar{n}_{\rm c})}{2N}\right)^{2s}}\left(e^{j2\pi \frac{l}{N}n_{\rm c}}-e^{j2\pi \frac{kM}{N}n_{\rm c}}\right)\left(e^{-j2\pi \frac{\bar{l}}{N}\bar{n}_{\rm c}}-e^{-j2\pi \frac{kM}{N}\bar{n}_{\rm c}}\right)}}  \nonumber \\
& = P_{\rm h} \sum\limits^{\infty}_{s=0}{\frac{(-1)^s}{(s!)^2}\left(\frac{\pi k_{D}}{N}\right)^{2s}} \sum\limits_{\bar{n}_{\rm c}\in \mathcal{N}}{\left(e^{-j2\pi \frac{\bar{l}}{N}\bar{n}_{\rm c}}-e^{-j2\pi \frac{kM}{N}\bar{n}_{\rm c}}\right)\sum\limits_{n_{\rm c}\in \mathcal{N}}{(n_{\rm c}-\bar{n}_{\rm c})^{2s}\left(e^{j2\pi \frac{l}{N}n_{\rm c}}-e^{j2\pi \frac{kM}{N}n_{\rm c}}\right)}}  \nonumber \\ 
& = R_{\rm H}(l, \bar{l}, kM),  
\end{align}
where $J_0(\cdot)$ is the zeroth order Bessel function of the first kind and $n_{\rm c}$ (or $\bar{n}_{\rm c}$) is the index of the channel path. It is noted from \eqref{Eqn22} that the large distance between $l$ (or $\bar{l}$) and $kM$ will increase the differences of the exponential functions. When the distance between $kM$ and $l$ is smaller than or equal to the coherence bandwidth, the differences of the exponential functions are small, that is $H(l)$ is close to $H(kM)$. Thus, the result of \eqref{Eqn22} is small. On the contrary, when the distance between $kM$ and $l$ exceeds the coherence bandwidth, the increased difference $H(l)-H(kM)$ enlarges $R_{\rm H}(l, \bar{l}, kM)$. On the other hand, with the reduced maximum channel time delay, that is the increased coherence bandwidth, the difference of $n_{\rm c}-\bar{n}_{\rm c}$ in \eqref{Eqn22} becomes small and $R_{\rm H}(l, \bar{l}, kM)$ also becomes small.

From \eqref{Eqn5}, we can obtain
\begin{align}
E\left\{X^*(l)X(\bar{l})\right\}&=E\left\{\sum\limits^{K-1}_{\bar{k}=0}{\sum\limits^{M-1}_{\bar{m}=0}{G^{*}_{\bar{k},\bar{m}}(l)d^{*}_{\bar{k},\bar{m}}}}\sum\limits^{K-1}_{\tilde{k}=0}{\sum\limits^{M-1}_{\tilde{m}=0}{G_{\tilde{k},\tilde{m}}(\bar{l})d_{\tilde{k},\tilde{m}}}}\right\} \nonumber  \\
&=\sum\limits^{K-1}_{\bar{k}=0} \sum\limits^{M-1}_{\bar{m}=0} G^{*}_{\bar{k},\bar{m}}(l)G_{\bar{k},\bar{m}}(\bar{l})E\left\{\left|d_{\bar{k},\bar{m}}\right|^2\right\}  \nonumber  \\
&=\sum\limits^{K-1}_{\bar{k}=0} \sum\limits^{M-1}_{\bar{m}=0} G^{*}_{\bar{k},\bar{m}}(l)G_{\bar{k},\bar{m}}(\bar{l}).  \nonumber  
\end{align}
Moreover, according to the local property of $G_{\bar{k},\bar{m}}(l)$, we can get $G^{*}_{\bar{k},\bar{m}}(l)G_{\bar{k},\bar{m}}((\bar{l})_N)=0$ when $|l-\bar{l}|>\tau$. 

\begin{figure}[h]
\centering
\includegraphics[width=5in]{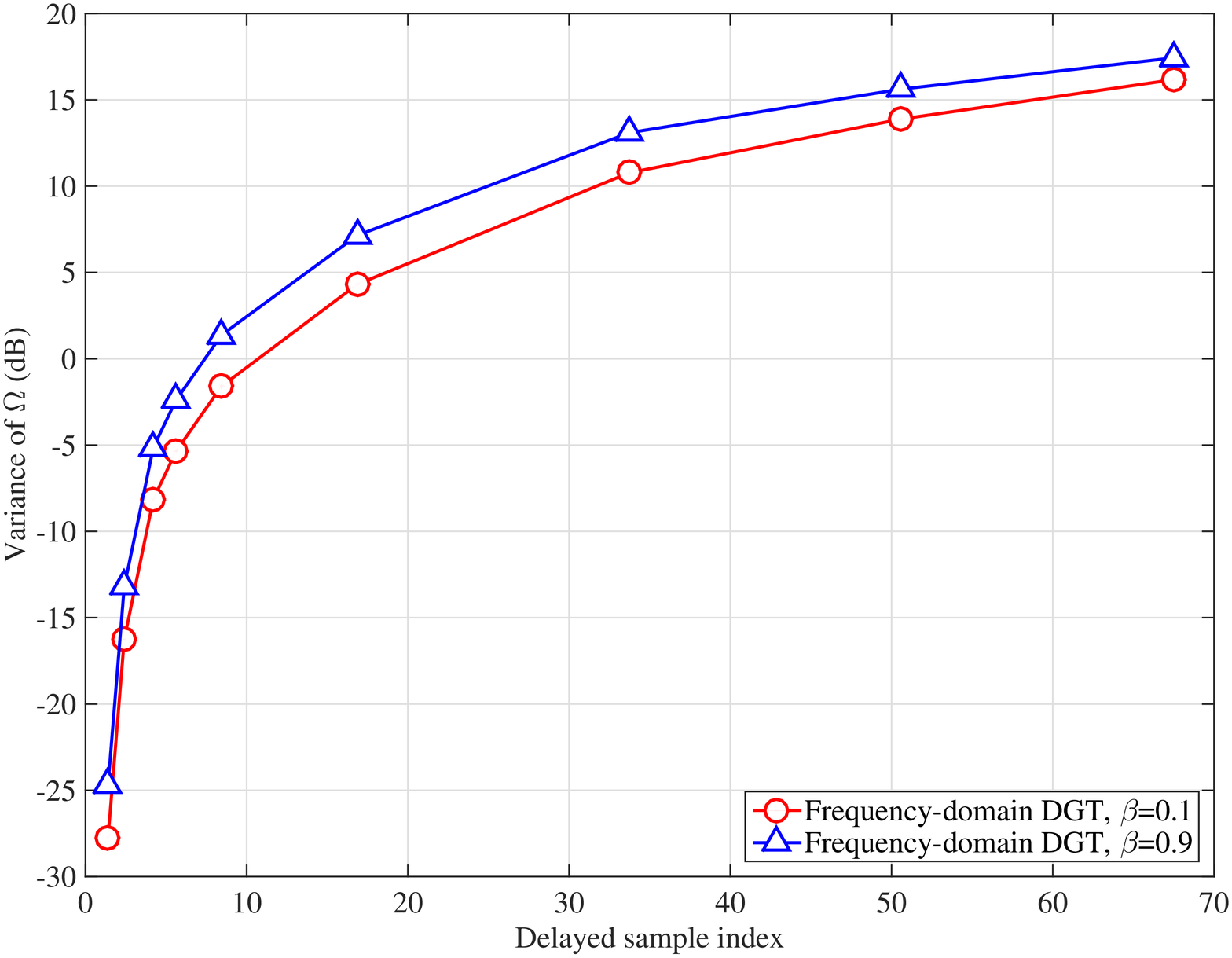}
 \caption{ Variances of $\Omega_{k,m}$ with different channel delays and roll-off factors under the normalized energy of ${\Gamma}_{k,m}(l)$ in the 9-path Rayleigh fading channel with the maximum channel delay $2.51\times10^{-6}$s.}
 \label{fig:fig3}
\end{figure}

\begin{figure}[h]
\centering
\includegraphics[width=5in]{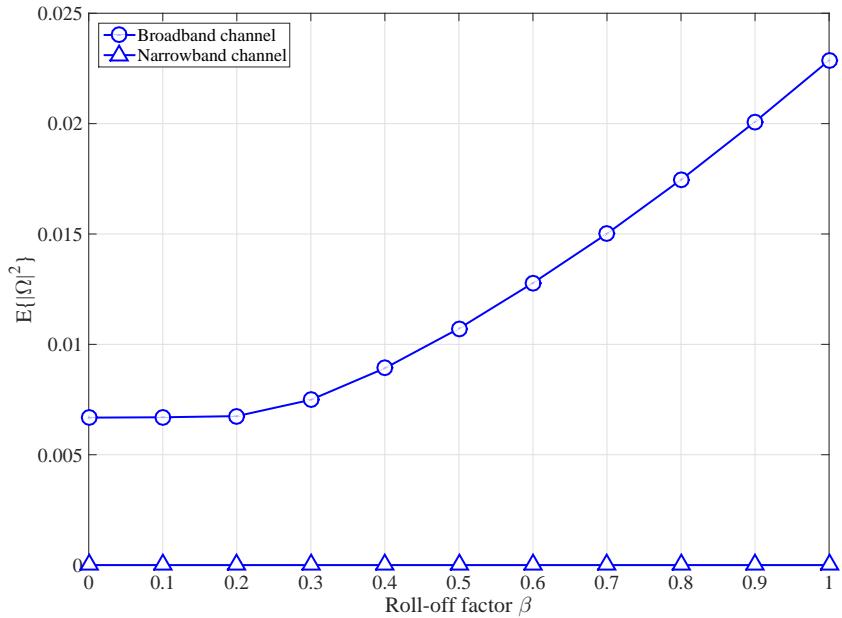}
 \caption{ Variances of $\Omega_{k,m}$ with different roll-off factors in a broadband channel and a narrowband channel.}
 \label{fig:fig8}
\end{figure}

Thus, the variance of $\Omega_{k,m}$ is further given by
\begin{align}
 \label{Eqn23}
E\left\{\Omega^{*}_{k,m}\Omega_{k,m}\right\} & =\frac{1}{N^2}\sum\limits^{N-1}_{l=0} \sum\limits^{l+\tau}_{\bar{l}=l-\tau}R_{\rm H}(l, \bar{l}, kM) \nonumber \\
& \quad   \cdot \sum\limits^{K-1}_{\bar{k}=0} \sum\limits^{M-1}_{\bar{m}=0} G^{*}_{\bar{k},\bar{m}}(l)G_{\bar{k},\bar{m}}((\bar{l})_N){\Gamma}_{k,m}(l){\Gamma}^*_{k,m}((\bar{l})_N).    
\end{align}
Eq. \eqref{Eqn23} denotes that the variances of $\Omega_{k,m}$ is influenced by $R_{\rm H}(l, \bar{l}, kM)$ and the product of $G_{k,m}(l)$ and ${\Gamma}_{k,m}(l)$, where $R_{\rm H}(l, \bar{l}, kM)$ decreases with the increase of the channel coherence and the product of $G_{k,m}(l)$ and ${\Gamma}_{k,m}(l)$ decreases with the decrease of the roll-off factor. Fig. \ref{fig:fig3} compares the variances of $\Omega_{k,m}$ with different maximum channel delays. It is shown that when the number of delayed signal samples equals to 1, the maximum channel delay is far smaller than the length, \emph{N}, of a GFDM symbol, and thus $n_{\rm c}-\bar{n}_{\rm c}$ approaches zero. The result is that the  summation of $R_{\rm H}(l, \bar{l}, kM)$ is close to zero and the variance of $\Omega_{k,m}$ approaches zeros. Obviously, in AWGN channel, the whole band is completely flat without channel delay, that is $n_{\rm c}-\bar{n}_{\rm c}=0$ ( or $H(l)=1$), we obtain $R_{\rm H}(l, \bar{l}, kM)=0$ and $\Omega_{k,m}=0$, similar to the narrowband channel shown in Fig. \ref{fig:fig8}. 
Thus, the maximum channel delay, related to the reciprocal of the coherence bandwidth, is the key factor of the variance of $\Omega_{k,m}$. On the other hand, with the increased roll-off factor of $G(l)$, the frequency-domain DGT enlarges the variance of $\Omega_{k,m}$, as shown in Fig. \ref{fig:fig8}, due to the decreased time-frequency localization of ${\Gamma}_{k,m}(l)$ and $G_{k,m}(l)$. When the roll-off factor is $\beta=0$, the synthesis window $G(l)$ becomes the rectangular window and its support length $2\tau+1$ becomes $M$:
\begin{equation}
G((l)_N)=\begin{cases}
    1,  & l \in [-\frac{M}{2}, \frac{M}{2}-1] \; \text{for even \emph{M}}, \; l \in [-\frac{M-1}{2}, \frac{M-1}{2}] \; \text{for odd \emph{M}}, \\
   0,   & \text{otherwise},
\end{cases} \nonumber 
\end{equation}
 and $\Gamma(l)$ is also the same rectangular window as $G(l)$ \cite{Ref4}. In this case, $X(l)$ in \eqref{Eqn5} and \eqref{Eqn7} becomes \emph{K} many \emph{M}-point DFTs:
\begin{equation}
X(l)=\sum\limits^{K-1}_{k=0}G((l+kM)_N) \sum\limits^{M-1}_{m=0}{d_{k,m}e^{-j2\pi\frac{m}{M}l}}= \sum\limits^{M-1}_{m=0}{d_{k,m}e^{-j2\pi\frac{m}{M}l}}\nonumber 
\end{equation}
for $l\in\mathcal{V}_{k}=[N-(k+\frac{1}{2})M+\eta, \; N-1-kM]\cup[(N-kM)_N, \; (N-1-(k-\frac{1}{2})M+\eta)_N]$, where when \emph{M} is even, $\eta=0$ and when \emph{M} is odd, $\eta=\frac{1}{2}$. Similarly, $Y_{k,m}$ in \eqref{Eqn18} becomes the \emph{M}-point IDFT
\begin{align}
Y_{k,m}&=\frac{1}{M}\sum\limits_{l\in\mathcal{V}_{k}}{Y(l)e^{j2\pi\frac{m}{M}l}}.  \nonumber 
\end{align}
$\Omega_{k,m}$ in \eqref{Eqn19} becomes
\begin{align}
\Omega_{k,m}=\frac{1}{M}\sum\limits_{l\in\mathcal{V}_{k}}{(H(l)-H(kM))X(l)e^{j2\pi\frac{m}{M}l}}. \nonumber
\end{align}


The variance of $\Psi_{k,m}$ in \eqref{Eqn20} is 
\begin{align}
E\left\{\Psi^{\rm *}_{k,m}\Psi_{k,m}\right\}&=E\left\{\left\|\left({\mathbf{\Gamma}}^{*}_{k,m}\right)^{\rm T}\mathbf{W}\right\|^2_2\right\}=\text{Tr}\left\{E\left\{{\mathbf{\Gamma}}^{\rm H}_{k,m}\mathbf{W}\mathbf{W}^{\rm H}{\mathbf{\Gamma}}_{k,m}\right\}\right\} \nonumber \\ 
&=\text{Tr}\left\{{\mathbf{\Gamma}}^{\rm H}_{k,m}E\left\{\mathbf{W}\mathbf{W}^{\rm H}\right\}{\mathbf{\Gamma}}_{k,m}\right\}  \nonumber \\
&=N\sigma^2\text{Tr}\left\{{\mathbf{\Gamma}}^{\rm H}_{k,m}{\mathbf{\Gamma}}_{k,m}\right\} =N\sigma^2\|\mathbf{\Gamma}\|^2_2, \nonumber
\end{align}
which denotes that the variance of $\Psi_{k,m}$ is unaffected by the frequency-domain DGT, but this noise will be colored now and may not be white anymore, where ${\mathbf{\Gamma}}_{k,m} \triangleq \frac{1}{N}[{\Gamma}_{k,m}(0), {\Gamma}_{k,m}(1), \ldots, {\Gamma}_{k,m}(N-1)]^{\rm T}$, $\mathbf{\Gamma} \triangleq \frac{1}{N}[\Gamma(0), \Gamma(1), \ldots, \Gamma(N-1)]^{\rm T}$ and $\mathbf{W} \triangleq [W(0), W(1), \ldots, W(N-1)]^{\rm T}$.



However, the frequency-domain DGT in the whole band still causes high complexity. Firstly, to get the received GFDM signal $Y(l)$, \emph{MK}-point FFT is required with $\frac{MK}{2}\log_2(MK)$ complex multiplications. Then, 
for the frequency-domain DGT in \eqref{Eqn18}, the number of complex multiplications required for \emph{K} many \emph{MK}-point circular convolutions between $Y(l)$ and $\Gamma(l)$ is $MK^2$. After that, based on the DFT-based DGT \cite{Ref4}, the frequency-domain DGT in the whole band of length \emph{MK} can be implemented by \emph{MK}-point FFT. Lastly, 
for detecting the data in \eqref{Eqn21}, $JMK$ complex multiplications are required from $H(kM)d_{k,m}$ and modulus in \eqref{Eqn21}. Thus, for a large \emph{M} or \emph{K}, the complexity $MK\log_2(MK)+MK^2+2JMK$ of the frequency-domain DGT receiver is high. In order to further reduce the complexity of the frequency-domain DGT in \eqref{Eqn18} at the receiver, the frequency-domain DGT in the local subbands is proposed below.

\subsection{Frequency-Domain Local DGT and A Fast Receiver}

\subsubsection{Frequency-Domain Local DGT}

Similar to the running window processing in time domain in \cite{Ref4, Ref13}, a signal $Y(l)$ with a localized analysis window $\Gamma(l)$ in the frequency domain called frequency-domain local DGT (LDGT) can be defined below. The LDGT of $Y(l)$ to get the (\emph{k}, \emph{m})-th data $d_{k,m}$ in the subband $[kM-L, kM+L]$ is defined by
\begin{equation}
\label{Eqn30}
\tilde{d}_{k,m}=\frac{1}{N}\sum\limits^{kM+L}_{l=kM-L}{Y(l)\tilde{\Gamma}^{*}_{k,m}(l)},
\end{equation}
where 
\begin{equation}
\label{Eqn39}
\tilde{\Gamma}_{k,m}(l)=\tilde{\Gamma}((l+kM)_N)e^{-j2\pi\frac{m}{M}l},
\end{equation}
which are the time and frequency shifts of an analysis window $\tilde{\Gamma}(l)$ for $l\in [0, L]\cup [N-L, N-1]$ and 2\emph{L}+1 is the support length of the analysis window $\tilde{\Gamma}(l)$. Note that an analysis window function usually has lowpass property, the non-zero elements of $\tilde{\Gamma}(l)$ are 
$\tilde{\Gamma}(0),\ldots, \tilde{\Gamma}(L), \; \tilde{\Gamma}(N-L),\ldots, \tilde{\Gamma}(N-1)$. The biorthogonality relationship between the synthesis window and the analysis window becomes  
\begin{align}
\label{Eqn31}
& \frac{1}{N}\sum\limits^{L}_{l=-L} {G(((l)_N+mM)_N)e^{j2\pi \frac{k}{M}(l)_N}\tilde{\Gamma}^{*}((l)_N)}=\delta(k)\delta(m) \nonumber \\  
&\qquad 0\leqslant k\leqslant M-1, \; \text{and} \;  0\leqslant m\leqslant K-1,
\end{align}
for $\tilde{d}_{k,m}=d_{k,m}$, $k=0,\ldots,K-1$, $m=0,\ldots,M-1$. Clearly when the synthesis window $G(l)$ is given, the local analysis window $\tilde{\Gamma}(l)$ can be solved from \eqref{Eqn31} if \eqref{Eqn31} has solutions. 

By rearranging \eqref{Eqn31} into a matrix vector form and deleting the all-zero rows, \eqref{Eqn31} becomes 
\begin{equation}
\label{Eqn36}
\mathbf{B} \tilde{\mathbf{\Gamma}}^{*}=\tilde{\mathbf{e}}_1,
\end{equation}
where $\mathbf{B}$ is a $(2\alpha -1 ) M \times (2L+1)$ matrix with $\alpha=\lceil \frac{L+\tau+1}{M} \rceil$, and $2 \tau +1$ is the non-zero length of the synthesis window $G(l)$, $(2 \alpha -1)M$ and $N-(2 \alpha -1)M$, respectively, denote the number of all nonzero rows and the number of all-zero rows in \eqref{Eqn31}, and thus the $(k+1+(m)_{2\alpha-1}M, l+L+1)$th element of $\mathbf{B}$ is 
\begin{equation}
\label{Eqn75}
G(((l)_N+(m)_{K}M)_N)e^{j2\pi\frac{k}{M}(l)_N}=G(((l)_N+(m)_{K}M)_N)e^{-j2\pi\frac{M-k}{M}(l)_N}
\end{equation}
 for $k=0, \ldots, M-1$, $m\in[-\alpha+1, \alpha-1]$, and $l\in[-L, L]$,
$\tilde{\mathbf{\Gamma}} \triangleq \frac{1}{N}[ \tilde{\Gamma}(N-L), \ldots, \tilde{\Gamma}(N-1), \tilde{\Gamma}(0), \ldots, \tilde{\Gamma}(L)]^{\rm T}$, and $\tilde{\mathbf{e}}_1=[1,0,\ldots,0]^{\rm T}$ is a $(2\alpha -1 ) M\times 1$ vector with its first element equal to 1. 

The support length of $G(l)$ always satisfies $2\tau+1\geqslant M$, as an example, for the RC window shown in Fig. \ref{fig:fig1}, where $M=2\bar{\beta}\tau$ for an even \emph{M} and $M=2\bar{\beta}\tau+1$ for an odd \emph{M}. Since $0\leqslant \beta \leqslant 1$, we can obtain $2\tau+1\geqslant 2 \bar{\beta}\tau +1 \geqslant M$, where the equal sign can be obtained when $\beta=0$. As mentioned above, when $\beta=0$, the analysis window $\Gamma(l)$ becomes a rectangular window the same as $G(l)$ with the support length \emph{M}. In this case, the frequency-domain DGT becomes \emph{K} many \emph{M}-point DFTs. Then, $\mathbf{B}$ becomes an $M\times M$ DFT matrix and \eqref{Eqn36} has a unique solution. Thus, the data easily recovered by a DFT is unique and is also with the least-squared error. On the contrary, when $0<\beta \leqslant 1$, we can obtain $2\tau+1> 2 \bar{\beta}\tau +1 \geqslant M$. Then, we have $(2\alpha -1) M > 2(L+\tau+1)-M > 2L+1$ for $0<\beta \leqslant 1$, which means that there are more equations than unknowns in \eqref{Eqn36}. Therefore, in general, the system of linear equations  \eqref{Eqn36} does not have a solution. We next focus on the case of $0<\beta \leqslant 1$ in the GFDM system in the following. In this case, we find $\tilde{\mathbf{\Gamma}}$ in \eqref{Eqn36} by using the following least squares criterion: 
\begin{equation}
\label{Eqn57}
\tilde{\mathbf{\Gamma}}_{\rm opt}={\rm arg}\min\limits_{\tilde{\mathbf{\Gamma}}}\left\| \tilde{\mathbf{e}}_1- \mathbf{B}\tilde{\mathbf{\Gamma}}^{*}\right\|^2_2,
\end{equation}
whose solution is the pseudoinverse of ${\mathbf{B}}$, i.e.,
\begin{equation}
\label{Eqn51}
\tilde{\mathbf{\Gamma}}_{\rm opt}=((\mathbf{B}^{*})^{\rm H}\mathbf{B}^{*})^{-1}({\mathbf{B}}^{*})^{\rm H}\tilde{\mathbf{e}}^{*}_1=(\mathbf{B}^{\rm T}\mathbf{B}^{*})^{-1}\mathbf{B}^{\rm T}\tilde{\mathbf{e}}_1=(\mathbf{B}^{\rm T}\mathbf{B}^{*})^{-1}\tilde{\mathbf{G}}_0,
\end{equation}
where $\tilde{\mathbf{G}}_0 \triangleq [G(N-L), \ldots, G(N-1),G(0),\ldots, G(L)]^{\rm T}$ because $\mathbf{B}^{\rm T} \tilde{\mathbf{e}}_1=\tilde{\mathbf{G}}_0$.

In the following, we prove that the GFDM data $\tilde{d}_{k,m}$ demodulated by LDGT with the optimal solution $\tilde{\mathbf{\Gamma}}_{\rm opt}$ also have the least-squared error compared to the original GFDM data $d_{k,m}$ among all analysis window functions $\tilde{\Gamma}(l)$ of length 2\emph{L}+1 as above. Note that in this case it corresponds to the ideal channel.


In the GFDM system, according to \eqref{Eqn30}, the LDGT using the GFDM signal $X(l)$ can be rewritten in the matrix form as
 \begin{align}
 \label{Eqn76}
\tilde{\mathbf{d}}=\mathbf{A}\mathbf{d}, 
\end{align}
where $\tilde{\mathbf{d}}=[\tilde{d}_{0,0},\tilde{d}_{0,M-1}, \ldots, \tilde{d}_{0, 1},  \ldots, \tilde{d}_{K-1,0}, \tilde{d}_{K-1,M-1}, \ldots, \tilde{d}_{K-1,1}]^{\rm T}$,  ${\mathbf{d}} \triangleq [{d}_{0,0},{d}_{0,M-1}, \ldots, {d}_{0,1},  \ldots, \\ {d}_{K-1,0}, {d}_{K-1,M-1}, \ldots, {d}_{K-1,1}]^{\rm T}$, $\mathbf{A}=[\mathbf{A}_0 \;  \mathbf{A}_1 \; \cdots \;  \mathbf{A}_{K-1}]^{\rm T}$, $\mathbf{A}^{\rm T}_0=[\mathbf{A}_{0, 1} \; \mathbf{0}_{M \times (N-(2\alpha-1) M)} \; \mathbf{A}_{0, 2}]$ with $[\mathbf{A}_{0, 2} \; \mathbf{A}_{0, 1}]=\tilde{\mathbf{\Gamma}}^{\rm H}_{2L+1}\mathbf{B}^{\rm T}$ and the $M \times (N-(2\alpha-1) M)$ all-zero matrix $\mathbf{0}_{M \times (N-(2\alpha-1) M)}$, $\mathbf{A}^{\rm T}_k$ is the shifted version of $\mathbf{A}^{\rm T}_0$ where the (\emph{m}+1)th row of $\mathbf{A}^{\rm T}_k$ is the cyclic shift of the (\emph{m}+1)th row of $\mathbf{A}^{\rm T}_0$ by \emph{kM} for $m=0,\ldots,M-1$, the $(2L+1)\times M$ matrix $\tilde{\mathbf{\Gamma}}_{2L+1}$ is defined by
\begin{equation}
\label{Eqn74}
\tilde{\mathbf{\Gamma}}_{2L+1} \triangleq [\mathbf{\Phi}_0\tilde{\mathbf{\Gamma}}, \mathbf{\Phi}_{1}\tilde{\mathbf{\Gamma}}, \cdots, \mathbf{\Phi}_{M-1}\tilde{\mathbf{\Gamma}}]
\end{equation}
with 
\begin{equation}
\label{Eqn77}
\mathbf{\Phi}_{m}\triangleq {\rm diag}\{e^{-j2\pi\frac{m}{M}(N-L)}, \ldots, e^{-j2\pi\frac{m}{M}(N-1)}, 1, \ldots, e^{-j2\pi\frac{m}{M}L}\}.
\end{equation}
From the block-cyclic format of $\mathbf{A}$, we just need to study any sub-matrix $\mathbf{A}^{\rm T}_k$ in $\mathbf{A}$. By deleting the all-zero matrix $\mathbf{0}_{M \times (N-(2\alpha-1) M)}$, only $\tilde{\mathbf{\Gamma}}^{\rm H}_{2L+1}\mathbf{B}^{\rm T}$ in $\mathbf{A}^{\rm T}_k$ is left. In the following, we prove that $\tilde{\mathbf{\Gamma}}$ and its shifts $\mathbf{\Phi}_m\tilde{\mathbf{\Gamma}}$ in $\tilde{\mathbf{\Gamma}}_{2L+1}$ have the same optimal solution $\tilde{\mathbf{\Gamma}}_{\rm opt}$.

For simplicity, by replacing \emph{m} and \emph{k} with \emph{k} and $m^{\prime}$ in $\mathbf{B}$, corresponding to the data indices of $\mathbf{d}_{k,m}$ and $\tilde{\mathbf{d}}_{k,m}$, the $((M-m)_M+1, (M-m^{\prime})_M+1+(k)_{2\alpha-1}M)$th element of $\tilde{\mathbf{\Gamma}}^{\rm H}_{2L+1}\mathbf{B}^{\rm T}$ is given by  
\begin{equation}
\label{Eqn60}
 \frac{1}{N}\sum\limits^{L}_{l=-L}{G(((l)_N+(k)_{K}M)_N)\tilde{\Gamma}^{*}((l)_N)e^{j2\pi\frac{(l)_N}{M}(m-m^{\prime})}},
\end{equation}
where $k \in [-\alpha+1,\alpha-1]$, and $m, m^{\prime}=0, M-1, \ldots, 1$. To obtain $\tilde{d}_{k,m}=d_{k,m}$, the shifted analysis windows and the shifted synthesis windows in \eqref{Eqn60} should also satisfy the biorthogonality similar to \eqref{Eqn31}, i.e., \eqref{Eqn60} equals to $\delta(k)\delta(m-m^{\prime})$. Thus, based on \eqref{Eqn31} and \eqref{Eqn36}, \eqref{Eqn60} satisfying biorthogonality can be expressed as
\begin{equation}
\label{Eqn62}
\mathbf{B} \mathbf{\Phi}^{*}_{m}\tilde{\mathbf{\Gamma}}^{*}=\tilde{\mathbf{e}}_{m+1},
\end{equation}
where $\tilde{\mathbf{e}}_{m+1}=[0,\ldots,0,1,0,\ldots,0]^{\rm T}$ is a $(2\alpha-1)M\times 1$ vector with its $((M-m)_M+1)$th element equal to 1 for $m=0, M-1, \ldots, 1$. Since $\mathbf{B} \mathbf{\Phi}^{*}_{m}$ has the same numbers of rows and columns as $\mathbf{B}$, similar to \eqref{Eqn57}, we can also formulate 
\begin{equation}
\label{Eqn63}
\tilde{\mathbf{\Gamma}}_{\rm opt}={\rm arg}\min\limits_{\tilde{\mathbf{\Gamma}}}\left\| \tilde{\mathbf{e}}_{m+1}- \mathbf{B}\mathbf{\Phi}^{*}_{m}\tilde{\mathbf{\Gamma}}^{*}\right\|^2_2,
\end{equation}
with the optimal solution 
\begin{align}
\label{Eqn64}
\tilde{\mathbf{\Gamma}}_{\rm opt}
=((\mathbf{B}^{*}\mathbf{\Phi}_{m})^{\rm H}\mathbf{B}^{*}\mathbf{\Phi}_{m})^{-1}(\mathbf{B}^{*}\mathbf{\Phi}_{m})^{\rm H}\tilde{\mathbf{e}}^{*}_{m+1} 
=\mathbf{\Phi}^{*}_{m}(\mathbf{B}^{\rm T}\mathbf{B}^{*})^{-1}\mathbf{\Phi}_{m}\tilde{\mathbf{G}}_0 
= (\mathbf{\Phi}^{*}_{m}\mathbf{B}^{\rm T}\mathbf{B}^{*}\mathbf{\Phi}_{m})^{-1}\tilde{\mathbf{G}}_0,
\end{align} 
where the $(l+L+1, l^{\prime}+L+1)$th element of $\mathbf{\Phi}^{*}_{m}\mathbf{B}^{\rm T}\mathbf{B}^{*}\mathbf{\Phi}_{m}$ is expressed by
\begin{align}
\label{Eqn65}
&\sum\limits_{k\in\mathcal{K}}{\sum\limits^{M-1}_{m^{\prime}=0}{G(((l)_N+kM)_N)G^{*}(((l^{\prime})_N+kM)_N)e^{j\frac{2\pi}{M}(l-l^{\prime})(m-m^{\prime})}}} \nonumber \\
&=  \sum\limits_{k\in\mathcal{K}} {G(((l)_N+kM)_N)G^{*}(((l^{\prime})_N+kM)_N)e^{j\frac{2\pi}{M}(l-l^{\prime})m}}\sum\limits^{M-1}_{m^{\prime}=0}{e^{-j\frac{2\pi}{M}(l-l^{\prime})m^{\prime}}}  \nonumber \\
&=  \sum\limits_{k\in\mathcal{K}} { G(((l)_N+kM)_N)G^{*}(((l^{\prime})_N+kM)_N)e^{j\frac{2\pi}{M}(l-l^{\prime})m}} M \delta(l-l^{\prime}+pM)  \nonumber \\
&=M\sum\limits_{k\in\mathcal{K}} { G(((l)_N+kM)_N)G^{*}(((l^{\prime})_N+kM)_N) \delta(l-l^{\prime}+pM)},
\end{align}
for $\mathcal{K}=[0,\alpha-1]\cup[K-\alpha+1, K-1]$, $m=0, M-1,\ldots, 1$, $l,l^{\prime}\in[-L, L]$ and $p=0,\pm 1, \ldots, \pm \lfloor \frac{2L}{M} \rfloor$. From \eqref{Eqn65}, one can see that it is independent of \emph{m}. As a result, 
\begin{equation}
\label{Eqn67}
\mathbf{B}^{\rm T}\mathbf{B}^{*}=\mathbf{\Phi}^{*}_{0}\mathbf{B}^{\rm T}\mathbf{B}^{*}\mathbf{\Phi}_{0}=\mathbf{\Phi}^{*}_{m}\mathbf{B}^{\rm T}\mathbf{B}^{*}\mathbf{\Phi}_{m},
\end{equation}
which proves that the optimal solution with the least-squared error in \eqref{Eqn63} is identical to the optimal solution with the least-squared error in \eqref{Eqn57}.

Thus, using the optimal $\tilde{\mathbf{\Gamma}}_{\rm opt}$ for the LDGT of the GFDM signal, for any other $\tilde{\mathbf{\Gamma}}$ of length 2\emph{L}+1, according to \eqref{Eqn63}, we can obtain
\begin{equation}
\label{Eqn61}
\left\| \tilde{\mathbf{e}}_{m+1}- \mathbf{B}\mathbf{\Phi}^{*}_{m}\tilde{\mathbf{\Gamma}}^{*}_{\rm opt}\right\|^2_2 \leqslant  \left\| \tilde{\mathbf{e}}_{m+1}- \mathbf{B}\mathbf{\Phi}^{*}_{m}\tilde{\mathbf{\Gamma}}^{*}\right\|^2_2.
\end{equation}

We assume that ${\rm E}\{d^{*}_{k,m}d_{k,m}\}=1$. Since $d_{k,m}$ for all \emph{k} and \emph{m} are i.i.d., based on \eqref{Eqn61}, we have 
\begin{align}
\label{Eqn52}
{\rm E}\left\{ \left\| \mathbf{d}-\tilde{\mathbf{d}}_{\rm opt}\right\|^2_2\right\} 
&={\rm E}\left\{ \left\| \left(\mathbf{I}_{N}-\mathbf{A}_{\rm opt}\right)\mathbf{d}\right\|^2_2\right\} 
={\rm Tr}\left\{ \left(\mathbf{I}_{N}-\mathbf{A}_{\rm opt}\right)\left(\mathbf{I}_{N}-\mathbf{A}_{\rm opt}\right)^{\rm H}\right\} \nonumber \\
&=\sum\limits^{K-1}_{k=0}{\sum\limits^{M-1}_{m=0}{\left\|\tilde{\mathbf{e}}^{\rm T}_{m+1}-\tilde{\mathbf{\Gamma}}^{\rm H}_{\rm opt}\mathbf{\Phi}^{*}_{m}\mathbf{B}^{\rm T}\right\|^2_2}} 
\leqslant \sum\limits^{K-1}_{k=0}{\sum\limits^{M-1}_{m=0}{\left\| \tilde{\mathbf{e}}^{\rm T}_{m+1}-\tilde{\mathbf{\Gamma}}^{\rm H}\mathbf{\Phi}^{*}_{m}\mathbf{B}^{\rm T}\right\|^2_2}} \nonumber \\
& = {\rm Tr}\left\{ \left(\mathbf{I}_{N}-{\mathbf{A}}\right)\left(\mathbf{I}_{N}-{\mathbf{A}}\right)^{\rm H}\right\}  
={\rm E}\left\{ \left\| \mathbf{d}-\tilde{\mathbf{d}}\right\|^2_2\right\}  ,
\end{align}
where ${\mathbf{A}}_{\rm opt}$ can be obtained by replacing $\tilde{\mathbf{\Gamma}}$ in $\mathbf{A}$ with $\tilde{\mathbf{\Gamma}}_{\rm opt}$. It is concluded by \eqref{Eqn52} that the data $\tilde{d}_{k,m}$ demodulated by the LDGT with the optimal analysis window $\tilde{\mathbf{\Gamma}}_{\rm opt}$ has the  least-squared error compared to the original data $d_{k,m}$ among all analysis window functions $\tilde{\mathbf{\Gamma}}$ of length 2\emph{L}+1 as above. In this case, the channel is ideal.

\subsubsection{A Fast Receiver}

In the receiver for a broadband channel, similar to \eqref{Eqn18}-\eqref{Eqn20}, the LDGT for the received GFDM signal $Y(l)$ in \eqref{Eqn4} in the subband $[kM-L, kM+L]$ is given by
\begin{align}
\label{Eqn40}
\tilde{Y}_{k,m}&=\frac{1}{N}\sum\limits^{kM+L}_{l=kM-L}{Y((l)_N)\tilde{\Gamma}^{*}_{k,m}((l)_N)}  \nonumber\\
&=H(kM)d_{k,m}+\tilde{\Omega}_{k,m}+\tilde{\Psi}_{k,m},
\end{align}
where 
\begin{align}
\label{Eqn41}
\tilde{\Omega}_{k,m}&=\frac{1}{N}\sum\limits^{kM+L}_{l=kM-L}{H((l)_N)X((l)_N)\tilde{\Gamma}^{*}(((l)_N+kM)_N)e^{j2\pi\frac{m}{M}(l)_N}}  \nonumber \\
&\quad -\frac{1}{N}\sum\limits^{N-1}_{l=0}{H(kM)X(l){\Gamma}^{*}((l+kM)_N)e^{j2\pi\frac{m}{M}l}},
\end{align}
and 
\begin{equation}
\label{Eqn42}
\tilde{\Psi}_{k,m}=\frac{1}{N}\sum\limits^{kM+L}_{l=kM-L}{W((l)_N)\tilde{\Gamma}^{*}(((l)_N+kM)_N)e^{j2\pi\frac{m}{M}(l)_N}},
\end{equation}
where the local analysis window function $\tilde{\Gamma}(l)=\tilde{\Gamma}_{\rm opt}(l)$ obtained previously. Then, based on \eqref{Eqn40}, the (\emph{k}, \emph{m})-th symbol $d_{k, m}$ is detected by using $\tilde{Y}_{k,m}$ similar to \eqref{Eqn21}.

What is shown previously is that  the  local analysis window $\tilde{\Gamma}_{\rm opt}(l)$ is optimal in terms of the data recovery, when the  channel is ideal or narrowband.  One might ask what will happen for a broadband channel, i.e., what will happen if a different local analysis window function $\Gamma(l)$ of length 2\emph{L}+1 is used in \eqref{Eqn40}-\eqref{Eqn42}. An obvious local analysis window function is the truncated $\Gamma(l)$ obtained through \eqref{Eqn17} to the length of 2\emph{L}+1, i.e., the truncated frequency-domain DGT in \eqref{Eqn18} to the band $[kM-L, kM+L]$. In this way, we can also obtain a fast GFDM receiver with the same complexity as the LDGT, which can be expressed by
\begin{align}
\label{Eqn24}
\bar{Y}_{k,m}
&=\frac{1}{N}\sum\limits^{kM+L}_{{l}=kM-L}{Y(({l})_N){\Gamma}^{*}_{k,m}(({l})_N)} \nonumber \\
&=H(kM)d_{k,m}+\bar{\Omega}_{k,m}+\bar{\Psi}_{k,m},
\end{align}
where 
\begin{align}
\label{Eqn37}
\bar{\Omega}_{k,m}=&\frac{1}{N}\sum\limits^{kM+L}_{{l}=kM-L}{H(({l})_N)X(({l})_N)\Gamma^{*}((({l})_N+kM)_N)e^{j2\pi\frac{m}{M}(l)_N}} \nonumber \\
&-\dfrac{1}{N}\sum\limits^{N-1}_{l=0}{H(kM)X(l)\Gamma^{*}((l+kM)_N)e^{j2\pi\frac{m}{M}l}},
\end{align}
and
\begin{equation}
\label{Eqn38}
\bar{\Psi}_{k,m}=\frac{1}{N}\sum\limits^{kM+L}_{{l}=kM-L}{W(({l})_N)\Gamma^{*}((({l})_N+kM)_N)e^{j2\pi\frac{m}{M}(l)_N}}.
\end{equation}


We next give the optimal analysis window with the least-squared error in the LDGT for the received GFDM signal when the channel statistics is known. Firstly, based on \eqref{Eqn40}-\eqref{Eqn42}, the average-squared error between $H(kM)d_{k,m}$ and $\tilde{Y}_{k,m}$ is expressed by
\begin{align}
\label{Eqn53}
{\rm E}\left\{\left\| \tilde{\mathbf{H}}\mathbf{d} -\tilde{\mathbf{Y}} \right\|^2_2\right\} 
 &=  {\rm E}\left\{\left\| \tilde{\mathbf{H}}\mathbf{d}-\tilde{\mathbf{A}}\mathbf{d} -\mathbf{C}\mathbf{W} \right\|^2_2\right\}  \nonumber \\
&= {\rm E}\left\{{\rm Tr}\left\{\left(\tilde{\mathbf{H}} - \tilde{\mathbf{A}}\right)\mathbf{d}\mathbf{d}^{\rm H} \left(\tilde{\mathbf{H}} - \tilde{\mathbf{A}} \right)^{\rm H}\right\}\right\}+ {\rm E}\left\{{\rm Tr}\left\{ \mathbf{C}\mathbf{W}\mathbf{W}^{\rm H}\mathbf{C}^{\rm H} \right\} \right\}  \nonumber \\
&= {\rm E}\left\{{\rm Tr}\left\{\mathbf{d}\mathbf{d}^{\rm H} \left(\tilde{\mathbf{H}} - \tilde{\mathbf{A}} \right)^{\rm H}\left(\tilde{\mathbf{H}} - \tilde{\mathbf{A}}\right)\right\}\right\}+  {\rm Tr}\left\{ \mathbf{C}{\rm E}\left\{\mathbf{W}\mathbf{W}^{\rm H}\right\} \mathbf{C}^{\rm H} \right\}   \nonumber \\
&= {\rm Tr}\left\{ {\rm E}\left\{\mathbf{d}\mathbf{d}^{\rm H}\right\} {\rm E}\left\{ \left(\tilde{\mathbf{H}} - \tilde{\mathbf{A}} \right)^{\rm H}\left(\tilde{\mathbf{H}} - \tilde{\mathbf{A}}\right) \right\} \right\}
+ N\sigma^2 {\rm Tr}\left\{ \mathbf{C}\mathbf{I}_N\mathbf{C}^{\rm H} \right\}   \nonumber \\
&= {\rm Tr}\left\{ {\rm E}\left\{ \left(\tilde{\mathbf{H}} - \tilde{\mathbf{A}} \right) \left(\tilde{\mathbf{H}} - \tilde{\mathbf{A}}\right)^{\rm H} \right\} \right\}  + N\sigma^2 {\rm Tr}\left\{ \mathbf{C}\mathbf{C}^{\rm H} \right\}   
\end{align}
where the $N \times N$ channel matrix $\tilde{\mathbf{H}}$ is defined as
$$\tilde{\mathbf{H}} \triangleq \begin{bmatrix}
H(0)\mathbf{I}_M & {} & {} & {} \\
{} & H(M)\mathbf{I}_M & {} & {} \\
{} & {} & \ddots & {} \\
{} & {} & {} & H((K-1)M)\mathbf{I}_M \\
\end{bmatrix},$$
 $\tilde{\mathbf{A}}=[\tilde{\mathbf{A}}_0 \;  \tilde{\mathbf{A}}_1 \; \cdots \;  \tilde{\mathbf{A}}_{K-1}]^{\rm T}$, $\tilde{\mathbf{A}}^{\rm T}_0=[\tilde{\mathbf{A}}_{0, 1} \; \mathbf{0}_{M \times (N-(2\alpha-1) M)} \; \tilde{\mathbf{A}}_{0, 2}]$ with $[\tilde{\mathbf{A}}_{0, 2} \; \tilde{\mathbf{A}}_{0, 1}]=\tilde{\mathbf{\Gamma}}^{\rm H}_{2L+1}\mathbf{H}_{0, 2L+1}\mathbf{B}^{\rm T}$, $\tilde{\mathbf{A}}^{\rm T}_k$ is the shifted version of $\tilde{\tilde{\mathbf{A}}}^{\rm T}_0$ that is obtained from $\tilde{\mathbf{A}}^{\rm T}_0$ by replacing $\mathbf{H}_{0,2L+1}$ in $\tilde{\mathbf{A}}^{\rm T}_0$ with $\mathbf{H}_{k,2L+1}$, where the (\emph{m}+1)th row of $\tilde{\mathbf{A}}^{\rm T}_k$ is the cyclic shift of the (\emph{m}+1)th row of $\tilde{\tilde{\mathbf{A}}}^{\rm T}_0$ by \emph{kM} for $m=0,\ldots,M-1$, and the $(2L+1)\times (2L+1)$ diagonal channel matrix $\mathbf{H}_{k, 2L+1}$ is defined by
\begin{equation}
\label{Eqn78}
\mathbf{H}_{k, 2L+1} \triangleq {\rm diag}\left\{ H((kM-L)_N) \; \cdots \;H((kM-1)_N) \; H((kM)_N)\; \cdots \; H((kM+L)_N)\right\},
\end{equation}
$\tilde{\mathbf{\Gamma}}^{\rm H}_{2L+1}$ and $\mathbf{B}$ have been shown in \eqref{Eqn74} and \eqref{Eqn75}, respectively,
$\mathbf{C}=[{\mathbf{C}}_0 \;  {\mathbf{C}_1} \; \cdots \;  {\mathbf{C}}_{K-1}]^{\rm T}$, ${\mathbf{C}}^{\rm T}_0=[{\mathbf{C}}_{0, 1} \; \mathbf{0}_{M \times (N-(2\alpha-1) M)} \; {\mathbf{C}}_{0, 2}]$ with $[{\mathbf{C}}_{0, 2} \; {\mathbf{C}}_{0, 1}]=\tilde{\mathbf{\Gamma}}^{\rm H}_{2L+1}$,
 ${\mathbf{C}}^{\rm T}_k$ is shifted version of ${\mathbf{C}}^{\rm T}_0$ where the (\emph{m}+1)th row of ${\mathbf{C}}^{\rm T}_k$ is the cyclic shift of the (\emph{m}+1)th row of ${\mathbf{C}}^{\rm T}_0$ by \emph{kM} for $m=0,\ldots,M-1$. 
 
Considering the property of  \eqref{Eqn67}, we can further rewrite \eqref{Eqn53} as
\begin{align}
\label{Eqn72}
&{\rm Tr}\left\{ {\rm E}\left\{ \left(\tilde{\mathbf{H}} - \tilde{\mathbf{A}} \right) \left(\tilde{\mathbf{H}} - \tilde{\mathbf{A}}\right)^{\rm H} \right\} \right\}
+ N\sigma^2 {\rm Tr}\left\{ \mathbf{C}\mathbf{C}^{\rm H} \right\} \nonumber \\
&=\sum\limits^{K-1}_{k=0} {\rm Tr}\big\{{\rm E}\big\{\left( H(kM)\left[\mathbf{I}_M \; \mathbf{0}_{M\times(2\alpha-2)M}\right] - \tilde{\mathbf{\Gamma}}^{\rm H}_{2L+1} \mathbf{H}_{k, 2L+1}\mathbf{B}^{\rm T}\right) \nonumber \\
&\qquad\qquad \cdot \left(H(kM)\left[ \mathbf{I}_M \; \mathbf{0}_{M\times(2\alpha-2)M}\right] - \tilde{\mathbf{\Gamma}}^{\rm H}_{2L+1} \mathbf{H}_{k, 2L+1}\mathbf{B}^{\rm T}\right)^{\rm H} \big\}\big\} 
  +N\sigma^2\sum\limits^{K-1}_{k=0}{{\rm Tr}\left\{\tilde{\mathbf{\Gamma}}^{\rm H}_{2L+1}\tilde{\mathbf{\Gamma}}_{2L+1}\right\}}   \nonumber \\
&=\sum\limits^{K-1}_{k=0}{\sum\limits^{M-1}_{m=0}{{\rm Tr}\left\{ {\rm E}\left\{ \left(H(kM)\tilde{\mathbf{e}}^{\rm T}_{m+1}- \tilde{\mathbf{\Gamma}}^{\rm H}\mathbf{\Phi}^{*}_{m} \mathbf{H}_{k,2L+1} \mathbf{B}^{\rm T} \right) \left(H(kM)\tilde{\mathbf{e}}^{\rm T}_{m+1}- \tilde{\mathbf{\Gamma}}^{\rm H}\mathbf{\Phi}^{*}_{m} \mathbf{H}_{k,2L+1} \mathbf{B}^{\rm T} \right)^{\rm H}\right\}\right\}}}  \nonumber \\
 &\quad + N \sigma^2 \sum\limits^{K-1}_{k=0}{\sum\limits^{M-1}_{m=0}{\left\|  \tilde{\mathbf{\Gamma}}^{\rm H} \mathbf{\Phi}^{*}_{m} \right\|^2_2}}  \nonumber \\
&= \sum\limits^{K-1}_{k=0} \sum\limits^{M-1}_{m=0} \big({\rm E}\left\{ |H(kM)|^2 \right\}  - \tilde{\mathbf{G}}^{\rm H}_0{\rm E}\left\{H(kM)\mathbf{H}^{*}_{k,2L+1}\right\}\tilde{\mathbf{\Gamma}} - \tilde{\mathbf{\Gamma}}^{\rm H}{\rm E}\left\{\mathbf{H}_{k,2L+1} H^{*}(kM) \right\} \tilde{\mathbf{G}}_0  \nonumber \\
  &\qquad\qquad\quad + \tilde{\mathbf{\Gamma}}^{\rm H}{\rm E}\left\{\mathbf{\Phi}^{*}_{m}\mathbf{H}_{k,2L+1}\mathbf{B}^{\rm T}\mathbf{B}^{*}\mathbf{H}^{*}_{k,2L+1}\mathbf{\Phi}_{m}\right\}\tilde{\mathbf{\Gamma}}\big)   +   N^2 \sigma^2 \left\|\tilde{\mathbf{\Gamma}}\right\|^2_2  \nonumber \\
&= \sum\limits^{K-1}_{k=0} \sum\limits^{M-1}_{m=0} \big({\rm E}\left\{ |H(kM)|^2 \right\}  - \tilde{\mathbf{G}}^{\rm H}_0{\rm E}\left\{H(kM)\mathbf{H}^{*}_{k,2L+1}\right\}\tilde{\mathbf{\Gamma}} - \tilde{\mathbf{\Gamma}}^{\rm H}{\rm E}\left\{\mathbf{H}_{k,2L+1} H^{*}(kM) \right\} \tilde{\mathbf{G}}_0  \nonumber \\
  &\qquad\qquad\quad + \tilde{\mathbf{\Gamma}}^{\rm H}{\rm E}\left\{\mathbf{H}_{k,2L+1}\mathbf{B}^{\rm T}\mathbf{B}^{*}\mathbf{H}^{*}_{k,2L+1}\right\}\tilde{\mathbf{\Gamma}}\big)   +   N^2 \sigma^2 \left\|\tilde{\mathbf{\Gamma}}\right\|^2_2
  \end{align}

Under the constraint of the constant $\|\tilde{\mathbf{\Gamma}}\|^2_2$, to minimize the error in \eqref{Eqn72}, we just need to minimize the first term of \eqref{Eqn72}. Thus, for obtaining the analysis window $\tilde{\mathbf{\Gamma}}_{\rm opt}$ with the least-squared error, we formulate 
\begin{equation}
\label{Eqn80}
\tilde{\mathbf{\Gamma}}_{\rm opt}={\rm arg}\min\limits_{\tilde{\mathbf{\Gamma}}} \left\{ e \right\}. 
\end{equation}
where
\begin{align}
e=&{\rm E}\left\{ |H(kM)|^2 \right\}  - \tilde{\mathbf{G}}^{\rm H}_0{\rm E}\left\{H(kM)\mathbf{H}^{*}_{k,2L+1}\right\}\tilde{\mathbf{\Gamma}}  - \tilde{\mathbf{\Gamma}}^{\rm H}{\rm E}\left\{\mathbf{H}_{k,2L+1} H^{*}(kM) \right\} \tilde{\mathbf{G}}_0  \nonumber \\
& + \tilde{\mathbf{\Gamma}}^{\rm H}{\rm E}\left\{\mathbf{H}_{k,2L+1}\mathbf{B}^{\rm T}\mathbf{B}^{*}\mathbf{H}^{*}_{k,2L+1}\right\}\tilde{\mathbf{\Gamma}}. \nonumber
\end{align} 

By ${\partial e}/{\partial \tilde{\mathbf{\Gamma}}}=0$, we have
\begin{align}
\label{Eqn81}
{\rm E}\left\{\mathbf{H}_{k,2L+1}\mathbf{B}^{\rm T}\mathbf{B}^{*}\mathbf{H}^{*}_{k,2L+1}\right\}\tilde{\mathbf{\Gamma}}_{\rm opt} 
={\rm E}\left\{H^{*}(kM)\mathbf{H}_{k,2L+1}\right\}\tilde{\mathbf{G}}_0 .    
\end{align}

Therefore, the optimal solution is
\begin{equation}
\label{Eqn82}
\tilde{\mathbf{\Gamma}}_{\rm opt}=\left( {\rm E}\left\{\mathbf{H}_{k,2L+1}\mathbf{B}^{\rm T}\mathbf{B}^{*}\mathbf{H}^{*}_{k,2L+1}\right\}  \right)^{-1} {\rm E}\left\{H^{*}(kM)\mathbf{H}_{k,2L+1}\right\}\tilde{\mathbf{G}}_0 .    
\end{equation}

For simplicity, by replacing \emph{k} and \emph{m} of $\mathbf{B}$ in \eqref{Eqn75} with \emph{m} and \emph{k}, according to \eqref{Eqn22}, the $(l+L+1, l^{\prime}+L+1)$th element of ${\rm E} \{\mathbf{H}_{k,2L+1}\mathbf{B}^{\rm T}\mathbf{B}^{*}\mathbf{H}^{*}_{k,2L+1} \}$ in \eqref{Eqn82} is expressed by
\begin{align}
\label{Eqn83}
&\sum\limits_{k\in\mathcal{K}} \sum\limits^{M-1}_{m=0}  {\rm E}\left\{H((l+kM)_N)H^{*}((l^{\prime}+kM)_N)\right\}  G(((l)_N+kM)_N)G^{*}(((l^{\prime})_N+kM)_N)e^{-j\frac{2\pi}{M}(l-l^{\prime})m}   \nonumber \\
&=  \sum\limits_{k\in\mathcal{K}}{ {\rm E}\left\{H((l+kM)_N)H^{*}((l^{\prime}+kM)_N)\right\} G(((l)_N+kM)_N)G^{*}(((l^{\prime})_N+kM)_N) }  \sum\limits^{M-1}_{m^{\prime}=0}{e^{-j\frac{2\pi}{M}(l-l^{\prime})m}}  \nonumber \\
&=M\sum\limits_{k\in\mathcal{K}}  {\rm E}\left\{H((l+kM)_N)H^{*}((l^{\prime}+kM)_N)\right\}  G(((l)_N+kM)_N)G^{*}(((l^{\prime})_N+kM)_N) \delta(l-l^{\prime}+pM)  \nonumber \\
&= M\sum\limits_{k\in\mathcal{K}} G(((l)_N+kM)_N)G^{*}(((l^{\prime})_N+kM)_N) \delta(l-l^{\prime}+pM)  \nonumber \\
& \qquad\qquad \cdot \sum\limits_{{n}_{\rm c}\in \mathcal{N}}\sum\limits_{{n}^{\prime}_{\rm c}\in \mathcal{N}} P_{\rm h} J_0 \left(2\pi\frac{k_{D}}{N}(n^{\prime}_{\rm c}-{n}_{\rm c})\right) e^{-j\frac{2\pi}{N}((l+kM)_N n_c - (l^{\prime}+kM)_N n^{\prime}_c)}  \nonumber \\
& = P_{\rm h} M\sum\limits_{k\in\mathcal{K}} G(((l)_N+kM)_N)G^{*}(((l^{\prime})_N+kM)_N) \delta(l-l^{\prime}+pM)  \nonumber \\
& \qquad\qquad \cdot \sum\limits^{\infty}_{s=0}{\frac{(-1)^s}{(s!)^2}\left(\frac{\pi k_{D}}{N}\right)^{2s}}  \sum\limits_{{n}_{\rm c}\in \mathcal{N}}\sum\limits_{{n}^{\prime}_{\rm c}\in \mathcal{N}} (n^{\prime}_{\rm c}-{n}_{\rm c})^{2s} e^{-j\frac{2\pi}{N}((l+kM)_N n_c - (l^{\prime}+kM)_N n^{\prime}_c)}   
\end{align}
for $\mathcal{K}=[0,\alpha-1]\cup[K-\alpha+1, K-1]$, $l,l^{\prime}\in[-L, L]$, and $p=0,\pm 1, \ldots, \pm \lfloor \frac{2L}{M} \rfloor$. Meanwhile, according to \eqref{Eqn22}, the $(l+L+1)$th element of ${\rm E}\left\{H^{*}(kM)\mathbf{H}_{k,2L+1}\right\}\tilde{\mathbf{G}}_0$ in \eqref{Eqn82} is
\begin{align}
\label{Eqn84}
& {\rm E}\left\{ H^{*}(kM) H((l+kM)_N) \right\} G((l)_N)  \nonumber \\
&\quad = G((l)_N) \sum\limits_{{n}_{\rm c}\in \mathcal{N}}\sum\limits_{{n}^{\prime}_{\rm c}\in \mathcal{N}} P_{\rm h} J_0 \left(2\pi\frac{k_{D}}{N}(n^{\prime}_{\rm c}-{n}_{\rm c})\right) e^{-j\frac{2\pi}{N}((l+kM)_N n_c - kM n^{\prime}_c)} \nonumber \\
 &\quad =P_{\rm h} G((l)_N) \sum\limits^{\infty}_{s=0}{\frac{(-1)^s}{(s!)^2}\left(\frac{\pi k_{D}}{N}\right)^{2s}}  \sum\limits_{{n}_{\rm c}\in \mathcal{N}}\sum\limits_{{n}^{\prime}_{\rm c}\in \mathcal{N}} (n^{\prime}_{\rm c}-{n}_{\rm c})^{2s} e^{-j\frac{2\pi}{N}((l+kM)_N n_c - kM n^{\prime}_c)}   .
\end{align}
Eqs. \eqref{Eqn83} and \eqref{Eqn84} show that $\tilde{\mathbf{\Gamma}}_{\rm opt}$ in \eqref{Eqn82} is related to the synthesis window $G((l)_N)$ and its shifts and the channel covariance. When the channel is ideal, there is one channel delay, i.e., $n_c=n^{\prime}_c=0$. In this case, we have ${\rm E}\left\{ H((l+kM)_N)H^{*}((l^{\prime}+kM)_N) \right\}=P_{\rm h}$ and ${\rm E}\left\{ H^{*}(kM) H((l+kM)_N) \right\}=P_{\rm h}$. Thus, the optimal analysis window in \eqref{Eqn82} is the same as the optimal analysis window in \eqref{Eqn51}.

By decreasing the length of the analysis window $\tilde{\Gamma}(l)$ to $2L+1$, the complexity of the LDGT can be reduced compared to the frequency-domain DGT. After \emph{MK}-point FFT, the number of the complex multiplications of the convolutions between $Y(l)$ and $\tilde{\Gamma}(l)$ is reduced to $K(2L+1)$, and the number of multiplications based on FFT for the LDGT is reduced to $(L+1) \log_2(MK)$ in \eqref{Eqn40}. The same as the frequency-domain DGT receiver, the data detection in \eqref{Eqn21} after the LDGT is also used. Thus, for $L\ll MK $, the complexity $(\frac{MK}{2}+L+1) \log_2(MK)+K(2L+1)+2JMK$ of the LDGT receiver is lower than the complexity $MK\log_2(MK)+MK^2+2JMK$ of the frequency-domain DGT receiver.

\begin{figure} [hp]
     \centering
     \subfloat[]{
   \label{fig:fig6a}        
          \includegraphics[width=4.6in]{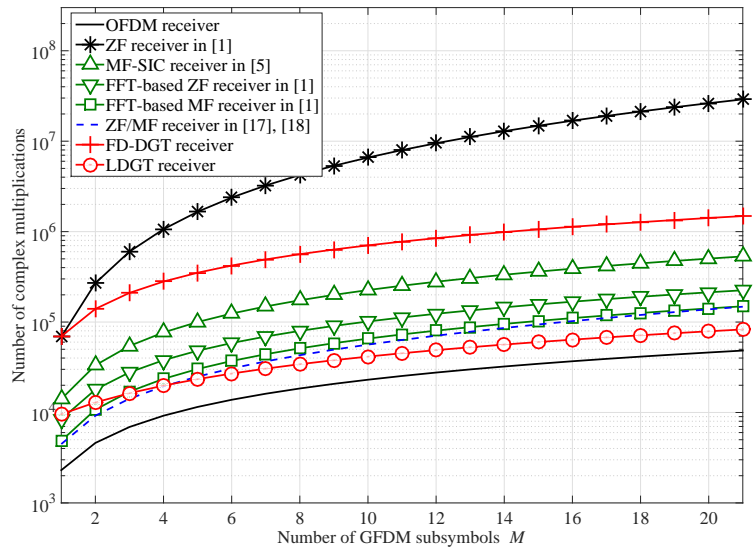}}
     \hspace{0.1\linewidth}  \\[20pt]  
     \subfloat[]{
          \label{fig:fig6b}        
          \includegraphics[width=4.6in]{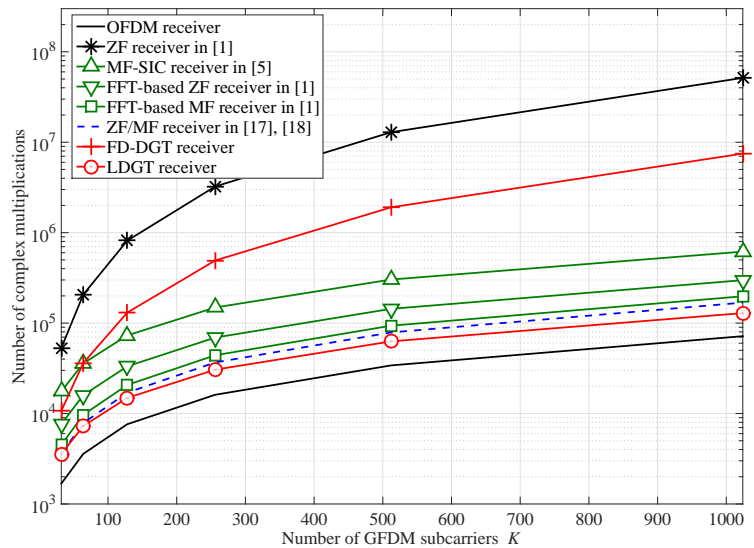}} 
 \caption{Computational complexity comparison of different GFDM receiver techniques in a broadband channel when $L=12$, $J=4$, and $I_0=8$. (a) $M\in[1, 21]$ and $K=256$; (b) $K\in\{32, 64,128,256,512,1024\}$ and $M=7$.}
     \label{fig:fig6}               
\end{figure}

Table \ref{table2} compares the complexities of several GFDM receivers in a broadband channel, where $I$ indicates the span of a receiver filter in the neighborhood of each subcarrier band in \cite{Ref1} and $I_0$ is the number of iterations in the SIC algorithm \cite{Ref10}. According to \cite{Ref1}, $I=2$ and $I=16$ are considered for the MF/MF-SIC and ZF receivers. Considering the channel equalization in OFDM, for fair complexity comparison, FDE is used as the channel equalization in the ZF receiver in \cite{Ref1}, the FFT-based ZF/MF receiver in \cite{Ref1}, the MF-SIC receiver in \cite{Ref10}, and the ZF/MF receiver for GFDM in \cite{Ref15, Ref16}. The FDE for the channel of length \emph{MK} in the GFDM receivers has $MK\log_2(MK)+MK$ complex multiplications caused by a pair of FFT and IFFT and ZF/MF. 
For simplicity, uncoded systems are considered here. Let \emph{J} be the size of the constellation $\mathcal{S}$. For $L\ll MK$, the LDGT in \eqref{Eqn40} can make a fast implementation of GFDM signal recovery. As shown in Fig. \ref{fig:fig6}, for small $M \leqslant 4$, the ZF/MF receiver for GFDM in \cite{Ref15, Ref16} has the lowest complexity, while the LDGT receiver has the complexity close to the ZF/MF receiver in \cite{Ref15, Ref16} and the FFT-based MF receiver in \cite{Ref1} and better than the FFT-based ZF receiver in \cite{Ref1}. On the contrary, when $M > 4$, the LDGT receiver has the lowest complexity among the GFDM receivers. 

\begin{table} [ht]
  \centering 
  \caption{Computational Complexities of Different GFDM Receiver Techniques in A Broadband Channel}
  \label{table2}
  \begin{tabular}{|c|c|}
\hline
\textbf{Technique} & \textbf{Number of complex multiplications} \\
 \hline
 OFDM receiver  &  $\frac{MK}{2}\log_2{K}+MK+JMK$ \\
  \hline
  ZF receiver in \cite{Ref1} & $(MK)^2+MK\log_2(MK)+MK+JMK$ \\  
  \hline
 MF-SIC receiver in \cite{Ref10}  &  $MK(\frac{3}{2}\log_2(MK)+\frac{1}{2}\log_2M+I+1+I_0(\log_2M+1+J))$ \\
   \hline
 FFT-based MF/ZF receiver in \cite{Ref1} & $MK (\frac{3}{2} \log_2(MK) + \frac{1}{2}\log_2{M} + I + 1+J)$    \\
  \hline
  ZF/MF receiver in \cite{Ref15, Ref16} & $\frac{MK}{2}(M+3\log_2{K}) +MK+JMK$ \\  
   \hline
 Frequency-domain DGT (FD-DGT) receiver & $MK\log_2{(MK)} + MK^2+2JMK$    \\
   \hline
  LDGT receiver & $(\frac{MK}{2}+L+1)\log_2{(MK)}+K(2L+1) + 2JMK$    \\
  \hline
\end{tabular}
\end{table}

\begin{figure}[ht]
\centering
\includegraphics[width=6in]{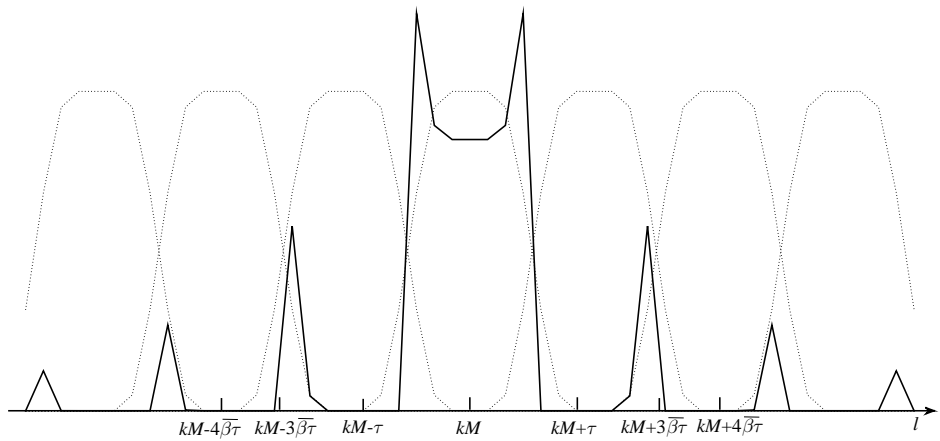}
\caption{The diagram of ${\Gamma}_{k,m}(l)$ of RC with $\beta=0.6$.} 
 \label{fig:fig2}
\end{figure}

After we study the above LDGT for the receiver of GFDM, it is clear that when the analysis window length 2\emph{L}+1 is increased, the receiver performance can be increased, while its complexity is increased as well. A simple way to trade-off the performance and the complexity of the LDGT receiver in choosing an analysis window length is as follows. First, we observe the whole band analysis window $\Gamma(l)$ obtained from the Wexler-Raz identity \eqref{Eqn17} to see where its concentration is as shown in Fig. \ref{fig:fig2}. Clearly, if one wants to truncate this function, one may want to see where its main energy is, for example,  use its main lobe or so, which can determine the truncated window length 2\emph{L}+1. This, then, can be used as the length in the LDGT as the local analysis window length. As we have proved before, using the optimal local analysis window function is always better than or at least equal to the truncated window function, the performance of the LDGT with the  obtained optimal local analysis window function will be good. 

\section{Simulation Results}

%


In the following simulations, the parameters are listed in Table \ref{table1}. The 9-path EVA channel model in 3GPP LTE is used, whose channel delay and channel power are [0, 30, 150, 310, 370, 710, 1090, 1730, 2510] ns and [0, -1.5, -1.4, -3.6, -0.6, -9.1, -7.0, -12.0, -16.9] dB, respectively.  

\begin{table} [ht]
  \centering 
  \caption{Simulation Parameters }
  \label{table1}
  \begin{tabular}{|c|c|}
\hline
\textbf{Parameters} & \textbf{Values} \\
 \hline
 Constellation modulation  &  QPSK and 16QAM \\
  Transmitter filter & RC \\
  Roll-off factor ($\beta$) & 0.1 and 0.9    \\
  Number of subcarriers (\emph{K}) & 256    \\
  Number of subsymbols (\emph{M}) & 7    \\
  Subcarrier interval & 15 KHz \\
  Sampling interval & 37.2 ns \\
  Carrier frequency & 2 GHz \\
Channel code & convolutional code \\
Code rate & 0.5 \\
  Maximum Doppler shift ($f_D$) & 100 Hz \\
  Length of CP in GFDM &  80\\
  Length of CP in OFDM & 80  \\
  Channel environment & multipath Rayleigh fading channel \\
\hline
\end{tabular}
\end{table}

\begin{figure} [hp]
     \centering
     \subfloat[]{
   \label{fig:fig4a}        
          \includegraphics[width=4.7in]{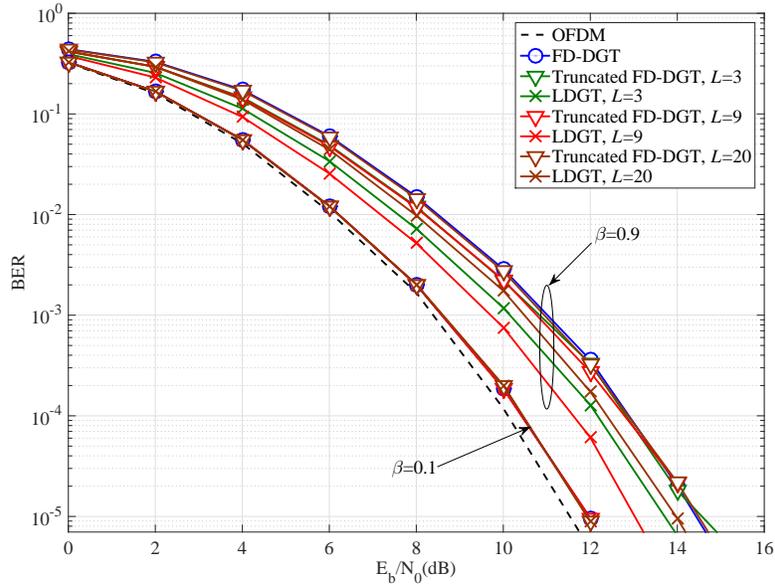}}
     \hspace{0.1\linewidth}  \\[20pt]  
     \subfloat[]{
          \label{fig:fig4b}        
          \includegraphics[width=4.7in]{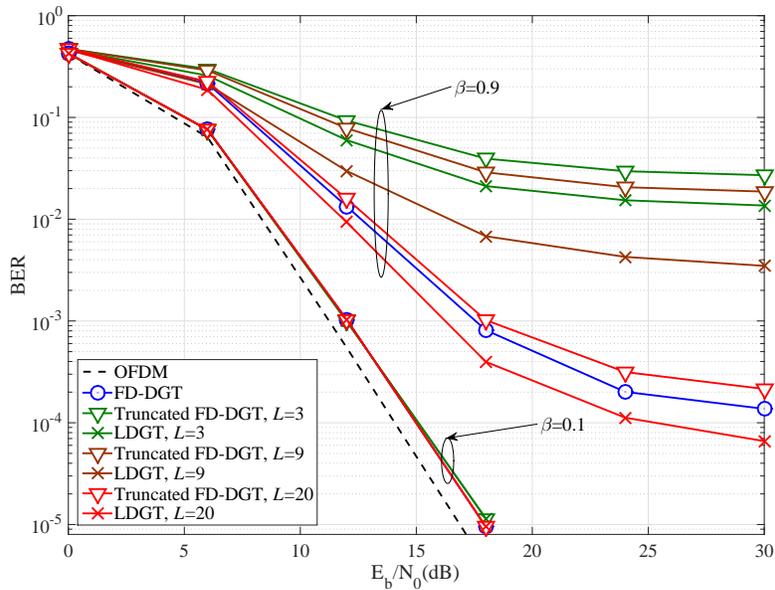}}  
 \caption{BERs of the GFDM signal processed by the frequency-domain DGT, the truncated frequency-domain DGT, and the LDGT with different lengths of the analysis windows and different roll-off factors of the synthesis windows in Rayleigh fading channels. (a) QPSK; (b) 16QAM.}
     \label{fig:fig4}               
\end{figure}

In Fig. \ref{fig:fig4}, the BER performances of the frequency-domain DGT, the truncated frequency-domain DGT and the LDGT with varying lengths of the analysis window and varying roll-off factors are depicted in Rayleigh fading channel. It is shown that the LDGT can obtain better BER performance than the truncated frequency-domain DGT, such as for $\beta=0.9$ and \emph{L}=9 in QPSK and $\beta=0.9$ and \emph{L}=20 in 16QAM. Compared to the frequency-domain DGT, the LDGT has the system performance degradation for the inaccurate $\tilde{\mathbf{\Gamma}}$, which is the analysis window in the local subband, obtained by the least squares criterion in \eqref{Eqn57}, but with the increased \emph{L}, the LDGT can obtain better BER performance than the frequency-domain DGT for the improved accuracy of  $\tilde{\mathbf{\Gamma}}$ and the removal of the part of the channel noise due to the local property of $\tilde{\mathbf{\Gamma}}$. For example, when $\beta=0.9$ and \emph{L}=9 in QPSK and $\beta=0.9$, \emph{L}=20 in 16QAM, the LDGT can obtain better BER performance than the frequency-domain DGT, while the truncated frequency-domain DGT cannot do. Meanwhile, the complexity of the LDGT in \eqref{Eqn40}, the same as the truncated frequency-domain DGT in \eqref{Eqn24}, is significantly reduced compared to the frequency-domain DGT in \eqref{Eqn18}, such as when $\beta=0.9$, \emph{L}=20 in 16QAM, the complexity reduction ratio is 85.5\%. Furthermore, with a small roll-off factor, both the LDGT and the truncated frequency-domain DGT can obtain the same BER performance as the frequency domain DGT in the whole band, such as $\beta=0.1$. It is concluded that compared to to the frequency-domain DGT in the whole band, the LDGT with a small length of the analysis window has significant complexity reduction while it can achieve a similar or better error performance.

\begin{figure}[htp]
\begin{center}
\includegraphics[width=4.7in]{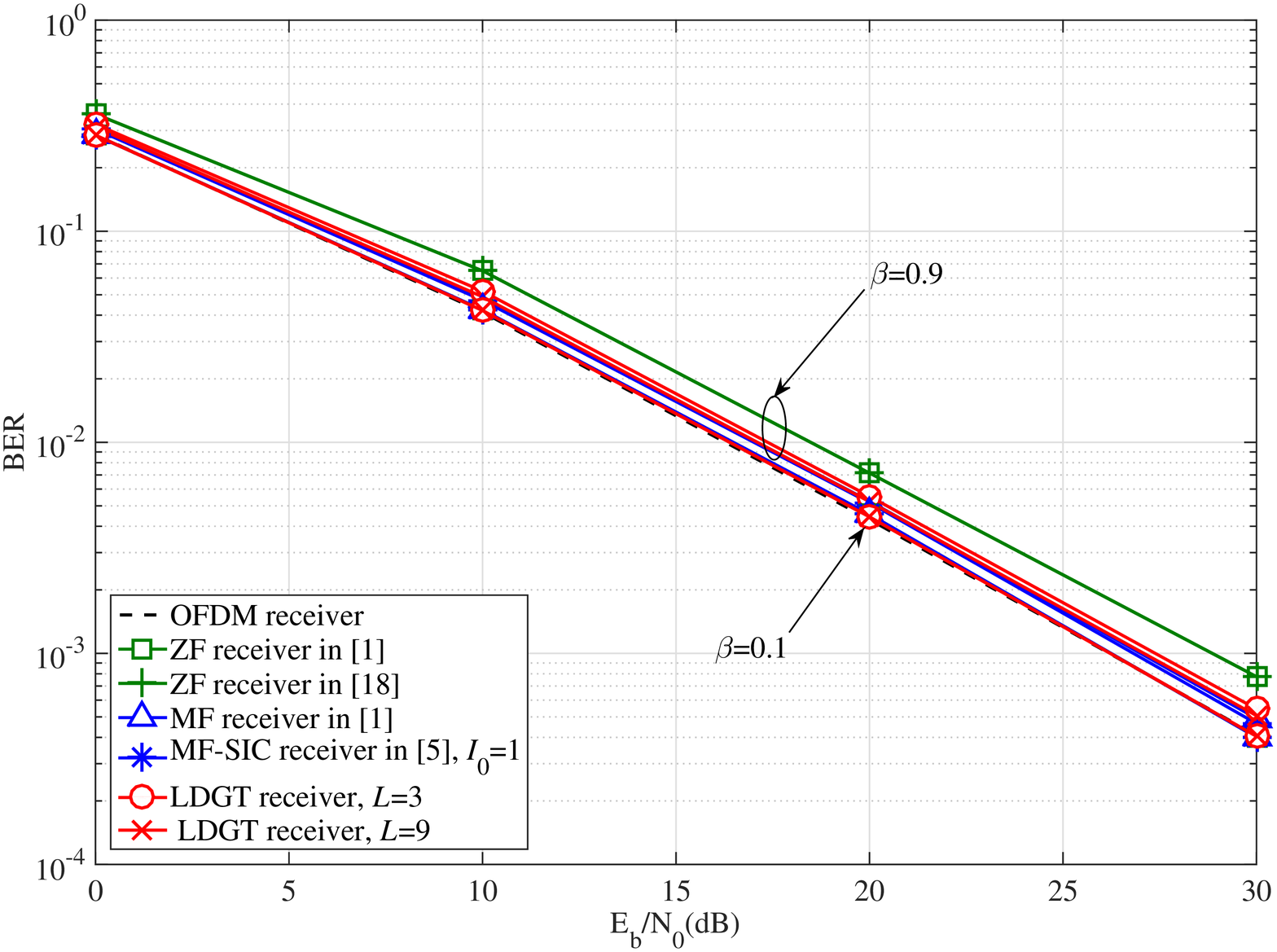}  
\caption{BER performance comparison among several detection methods for the GFDM signal in a narrowband channel with channel delay 37.2 ps and channel power 0 dB, where QPSK is adopted.}
\label{fig:fig7}
\end{center}
\end{figure}

\begin{figure}[htp]
\begin{center}
\includegraphics[width=4.7in]{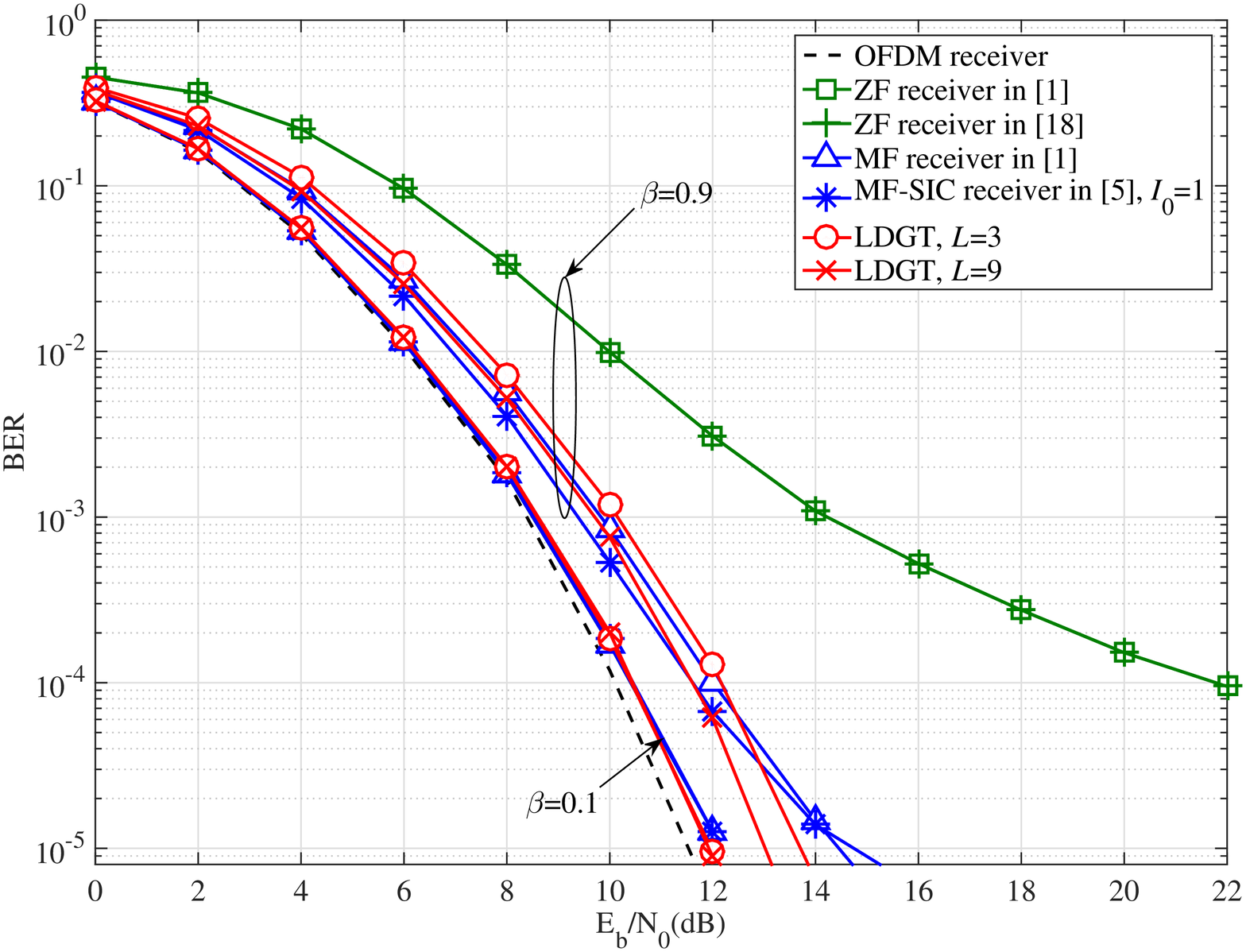}
\caption{BER performance comparison among several detection methods for the GFDM signal in 9-path Rayleigh fading channel, where QPSK is adopted.}
\label{fig:fig5}
\end{center}
\end{figure}

Figs. \ref{fig:fig7} and \ref{fig:fig5} compare the BER performances among the ZF receiver in \cite{Ref1}, the FFT-based MF receiver in \cite{Ref1}, the MF-SIC receiver in \cite{Ref10}, the ZF receiver in \cite{Ref16}, and the LDGT receiver in a narrowband channel and a broadband channel, respectively, where QPSK is adopted. Compared to the other GFDM receivers, the LDGT receiver shows the promising BER performance. The BER performance in the LDGT receiver can be significantly improved by a large \emph{L} or a small roll-off factor $\beta$. For example, let the parameter $L$ increase from $L=3$ to $L=9$ when $\beta=0.9$ in the broadband channel and  the performances are shown in Fig. \ref{fig:fig5}. In this case, the LDGT receiver can obtain the better BER performance than the ZF receiver in \cite{Ref1}, the ZF receiver in \cite{Ref16}, and the MF-SIC receiver with $I_0$=1. This is because our proposed LDGT receiver does not use a direct  channel equalization or the symbol-by-symbol detection in \eqref{Eqn21} to calculate the soft information of the channel decoder. However, before the calculation of the soft information, the other GFDM receivers in \cite{Ref1,Ref10,Ref16} still employ channel equalization before decoding. Without consideration of the complexity of the soft information calculation and the channel decoding, according to Table \ref{table2}, in the coded GFDM system with $\beta=0.9$, compared to the ZF receiver in \cite{Ref1}, the MF-SIC receiver in \cite{Ref10}, the ZF receiver in \cite{Ref16} and the FFT-based MF receiver in \cite{Ref1}, the complexity reduction ratios in LDGT receiver with $L=9$ are 99.6\%, 66.5\%, 50.4\%, and 60.3\%, respectively. Thus, the LDGT receiver has the lowest complexity while maintaining considerable BER performance in the broadband channel. 

\section{Conclusion}

In this paper, the transmitted GFDM signal was first considered as the IDGT in time domain and frequency domain, respectively. Then, for redcing the complexity caused by the channel equalization, we proposed the frequency-domain DGT for the received GFDM signal to simplify the GFDM signal recovery similar to OFDM. By analyzing the interference caused by the frequency-domain DGT, the channel with high coherence and a small roll-off factor of the synthesis widow can lead to small interference to the received signal. Based on the localized synthesis window in the frequency domain, the LDGT was proposed in the local band to further reduce the complexity of the frequency-domain DGT in the whole band. Although the truncation of the frequency-domain DGT can achieve the same complexity as the LDGT, we proved that the data demodulated by the LDGT with the optimal analysis window has the least-squared error in the ideal channel and the broadband channel compared to the truncated frequency-domain DGT. Simulation results showed that as the length of the optimal analysis window increases, the LDGT can obtain BER performance as good as the frequency-domain DGT, while having notable complexity reduction compared to other GFDM receivers.

\end{document}